%% Draft Mode
%\def\mydraft{label}           %turn on label
%\def\mydraft{no label}        %turn off label
%\def\drafttitle{minimal1.tex}   %display title on top-right

%%%%%%%%%%%%%%%%%%%%%%%%%%%%%%%%
%
%
% Vasilis Niarchos (Niels Bohr Institute)
%
%
%
%%%%%%%%%%%%%%%%%%%%%%%%%%%%%%%%

\input lanlmac

\input epsf

%Macro for figure
\newcount\figno
\figno=0
\def\fig#1#2#3{
\par\begingroup\parindent=0pt\leftskip=1cm\rightskip=1cm\parindent=0pt
\baselineskip=11pt
\global\advance\figno by 1
\midinsert
\epsfxsize=#3
\centerline{\epsfbox{#2}}
\vskip 12pt
\centerline{{\bf Fig. \the\figno:~~} #1}\par
\endinsert\endgroup\par
}
\def\figlabel#1{\xdef#1{\the\figno}}

%%%%%%%
% Older defs

\def\journal#1&#2(#3){\unskip, \sl #1\ \bf #2 \rm(19#3) }
\def\andjournal#1&#2(#3){\sl #1~\bf #2 \rm (19#3) }

\def\frac#1#2{{#1\over#2}}

\def\d{\partial}

\def\inbar{\,\vrule height1.5ex width.4pt depth0pt}
\def\IC{\relax\hbox{$\inbar\kern-.3em{\rm C}$}}
\def\IR{\relax{\rm I\kern-.18em R}}
\def\IP{\relax{\rm I\kern-.18em P}}
\def\IZ{\relax{\rm I\kern-.18em Z}}

%
%%%%%%%%%%%%%%%%%%%%%%%%%%%%%%%%%%%%
%

%
\catcode`\@=11
\def\slash#1{\mathord{\mathpalette\c@ncel{#1}}}
\overfullrule=0pt
\def\AA{{\cal A}}

\def\II{{\cal I}}

\def\LL{{\cal L}}
\def\MM{{\cal M}}
\def\NN{{\cal N}}
\def\OO{{\cal O}}

\def\SS{{\cal S}}

\def\VV{{\cal V}}

\def\ZZ{{\cal Z}}

\def\underrel#1\over#2{\mathrel{\mathop{\kern\z@#1}\limits_{#2}}}

\catcode`\@=12

%%%%%%%%%%%%%%%%%%%%%%%%%%%%%%%%%%%%%%%%%%%%%%%%%

%

\def\det{{\rm det}}
\def\tr{{\rm tr}}

\def \sinh{{\rm sinh}}
\def \cosh{{\rm cosh}}

\def\det{{\rm det}}

%%%%%%%%%%%%%%%%%%%%%%%%%%%%%%%%%%%%%%%%%%%%%%%%%
% new defs:

% Something to deal with sub-sub-sections

\def\unlockat{\catcode`\@=11}
\def\lockat{\catcode`\@=12}

\unlockat

%%% Something to deal with sub-sub-sections

\def\newsec#1{\global\advance\secno by1\message{(\the\secno. #1)}
\global\subsecno=0\global\subsubsecno=0\eqnres@t\noindent
{\bf\the\secno. #1}
\writetoca{{\secsym} {#1}}\par\nobreak\medskip\nobreak}
\global\newcount\subsecno \global\subsecno=0
\def\subsec#1{\global\advance\subsecno
by1\message{(\secsym\the\subsecno. #1)}
\ifnum\lastpenalty>9000\else\bigbreak\fi\global\subsubsecno=0
\noindent{\it\secsym\the\subsecno. #1}
\writetoca{\string\quad {\secsym\the\subsecno.} {#1}}
\par\nobreak\medskip\nobreak}
\global\newcount\subsubsecno \global\subsubsecno=0
\def\subsubsec#1{\global\advance\subsubsecno by1
\message{(\secsym\the\subsecno.\the\subsubsecno. #1)}
\ifnum\lastpenalty>9000\else\bigbreak\fi
\noindent\quad{\secsym\the\subsecno.\the\subsubsecno.}{#1}
\writetoca{\string\qquad{\secsym\the\subsecno.\the\subsubsecno.}{#1}}
\par\nobreak\medskip\nobreak}

\def\subsubseclab#1{\DefWarn#1\xdef
#1{\noexpand\hyperref{}{subsubsection}%
{\secsym\the\subsecno.\the\subsubsecno}%
{\secsym\the\subsecno.\the\subsubsecno}}%
\writedef{#1\leftbracket#1}\wrlabeL{#1=#1}}% Macros for boxes
\lockat

% End of Older Defs
%%%%%%%%%%%%%

%\def\tilde{\widetilde}
\newcount\figno
\figno=0
\def\fig#1#2#3{
\par\begingroup\parindent=0pt\leftskip=1cm\rightskip=1cm\parindent=0pt
\baselineskip=11pt
\global\advance\figno by 1
\midinsert
\epsfxsize=#3
\centerline{\epsfbox{#2}}
\vskip 12pt
{\bf Fig.\ \the\figno: } #1\par
\endinsert\endgroup\par
}
\def\figlabel#1{\xdef#1{\the\figno}}
\def\encadremath#1{\vbox{\hrule\hbox{\vrule\kern8pt\vbox{\kern8pt
\hbox{$\displaystyle #1$}\kern8pt}
\kern8pt\vrule}\hrule}}
%
%

%%% Paragraphs

%%% special math symbols
\font\cmss=cmss10
\font\cmsss=cmss10 at 7pt
\def\rlx{\relax\leavevmode}
\def\inbar{\vrule height1.5ex width.4pt depth0pt}
\def\IC{\relax\,\hbox{$\inbar\kern-.3em{\rm C}$}}
\def\IN{\relax{\rm I\kern-.18em N}}
\def\IP{\relax{\rm I\kern-.18em P}}
\def\ZZ{\rlx\leavevmode\ifmmode\mathchoice{\hbox{\cmss Z\kern-.4em Z}}
 {\hbox{\cmss Z\kern-.4em Z}}{\lower.9pt\hbox{\cmsss Z\kern-.36em Z}}
 {\lower1.2pt\hbox{\cmsss Z\kern-.36em Z}}\else{\cmss Z\kern-.4em
 Z}\fi}
%%% misc.
\def\IZ{\relax\ifmmode\mathchoice
{\hbox{\cmss Z\kern-.4em Z}}{\hbox{\cmss Z\kern-.4em Z}}
{\lower.9pt\hbox{\cmsss Z\kern-.4em Z}}
{\lower1.2pt\hbox{\cmsss Z\kern-.4em Z}}\else{\cmss Z\kern-.4em
Z}\fi}
%%% misc.
\def\IZ{\relax\ifmmode\mathchoice
{\hbox{\cmss Z\kern-.4em Z}}{\hbox{\cmss Z\kern-.4em Z}}
{\lower.9pt\hbox{\cmsss Z\kern-.4em Z}}
{\lower1.2pt\hbox{\cmsss Z\kern-.4em Z}}\else{\cmss Z\kern-.4em
Z}\fi}

\def\narrowplus{\kern -.04truein + \kern -.03truein}
\def\narrowminus{- \kern -.04truein}
\def\narrowminussub{\kern -.02truein - \kern -.01truein}

\def\IZ{\relax\ifmmode\mathchoice
{\hbox{\cmss Z\kern-.4em Z}}{\hbox{\cmss Z\kern-.4em Z}}
{\lower.9pt\hbox{\cmsss Z\kern-.4em Z}}
{\lower1.2pt\hbox{\cmsss Z\kern-.4em Z}}\else{\cmss Z\kern-.4em
Z}\fi}
\def\IB{\relax{\rm I\kern-.18em B}}
\def\IC{{\relax\hbox{$\inbar\kern-.3em{\rm C}$}}}
\def\ID{\relax{\rm I\kern-.18em D}}
\def\IE{\relax{\rm I\kern-.18em E}}
\def\IF{\relax{\rm I\kern-.18em F}}
\def\IG{\relax\hbox{$\inbar\kern-.3em{\rm G}$}}
\def\IGa{\relax\hbox{${\rm I}\kern-.18em\Gamma$}}
\def\IH{\relax{\rm I\kern-.18em H}}
\def\II{\relax{\rm I\kern-.18em I}}
\def\IK{\relax{\rm I\kern-.18em K}}
\def\IP{\relax{\rm I\kern-.18em P}}
%\def\IX{\relax{\rm X\kern-.01em X}}
%this doesn't work

\font\cmss=cmss10 \font\cmsss=cmss10 at 7pt
\def\IR{\relax{\rm I\kern-.18em R}}

%\def\mp{M_{\rm p}}

%

%
%       \eqn\label{a+b=c}       gives displayed equation, numbered
%                               consecutively within sections.
%     \eqnn and \eqna define labels in advance (of eqalign?)
%
\def\eqnn#1{\xdef #1{(\secsym\the\meqno)}\writedef{#1\leftbracket#1}%
\global\advance\meqno by1\wrlabeL#1}
\def\eqna#1{\xdef #1##1{\hbox{$(\secsym\the\meqno##1)$}}
\writedef{#1\numbersign1\leftbracket#1{\numbersign1}}%
\global\advance\meqno by1\wrlabeL{#1$\{\}$}}
\def\eqn#1#2{\xdef #1{(\secsym\the\meqno)}\writedef{#1\leftbracket#1}%
\global\advance\meqno by1$$#2\eqno#1\eqlabeL#1$$}

%%%%%%%%%%%%%%%%%%%%
% References

%\AharonyVK
\lref\AharonyVK{
O.~Aharony, B.~Fiol, D.~Kutasov and D.~A.~Sahakyan,
``Little string theory and heterotic/type II duality,''
Nucl.\ Phys.\ B {\bf 679}, 3 (2004)
[arXiv:hep-th/0310197].
%%CITATION = HEP-TH 0310197;%%
}

%\AharonyXN
\lref\AharonyXN{
  O.~Aharony, A.~Giveon and D.~Kutasov,
  ``LSZ in LST,''
  Nucl.\ Phys.\ B {\bf 691}, 3 (2004)
  [arXiv:hep-th/0404016].
  %%CITATION = HEP-TH 0404016;%%
}

%\GiveonZM
\lref\GiveonZM{
  A.~Giveon, D.~Kutasov and O.~Pelc,
  ``Holography for non-critical superstrings,''
  JHEP {\bf 9910}, 035 (1999)
  [arXiv:hep-th/9907178].
  %%CITATION = HEP-TH 9907178;%%
}

%\KutasovUF
\lref\KutasovUF{
  D.~Kutasov,
  ``Introduction to little string theory,''
%\href{http://www.slac.stanford.edu/spires/find/hep/www?irn=5015588}{SPIRES entry}
{\it Prepared for ICTP Spring School on Superstrings and Related
Matters, Trieste, Italy, 2-10 Apr 2001}
}

%\AharonyKS
\lref\AharonyKS{
  O.~Aharony,
  ``A brief review of 'little string theories',''
  Class.\ Quant.\ Grav.\  {\bf 17}, 929 (2000)
  [arXiv:hep-th/9911147].
  %%CITATION = HEP-TH 9911147;%%
}

%\GiveonPX
\lref\GiveonPX{
  A.~Giveon and D.~Kutasov,
  ``Little string theory in a double scaling limit,''
  JHEP {\bf 9910}, 034 (1999)
  [arXiv:hep-th/9909110].
  %%CITATION = HEP-TH 9909110;%%
}

%\ItzhakiTU
\lref\ItzhakiTU{
  N.~Itzhaki, D.~Kutasov and N.~Seiberg,
  ``I-brane dynamics,''
  arXiv:hep-th/0508025.
  %%CITATION = HEP-TH 0508025;%%
}

%\IsraelIR
\lref\IsraelIR{
  D.~Israel, C.~Kounnas, A.~Pakman and J.~Troost,
  ``The partition function of the supersymmetric two-dimensional black
hole
  and little string theory,''
  JHEP {\bf 0406}, 033 (2004)
  [arXiv:hep-th/0403237].
  %%CITATION = HEP-TH 0403237;%%
}

%\KiritsisMC
\lref\KiritsisMC{
  E.~Kiritsis,
  ``D-branes in standard model building, gravity and cosmology,''
  Fortsch.\ Phys.\  {\bf 52}, 200 (2004)
  [Phys.\ Rept.\  {\bf 421}, 105 (2005)]
  [arXiv:hep-th/0310001].
  %%CITATION = HEP-TH 0310001;%%
}

%\PolchinskiRR
\lref\PolchinskiRR{
  J.~Polchinski,
  ``String theory. Vol. 2: Superstring theory and beyond.''
%\href{http://www.slac.stanford.edu/spires/find/hep/www?irn=4634802}{SPIRES entry}
}

%\IsraelCD
\lref\IsraelCD{
  D.~Israel, C.~Kounnas, D.~Orlando and P.~M.~Petropoulos,
  ``Heterotic strings on homogeneous spaces,''
  Fortsch.\ Phys.\  {\bf 53}, 1030 (2005)
  [arXiv:hep-th/0412220].
  %%CITATION = HEP-TH 0412220;%%
}

%\IsraelVV
\lref\IsraelVV{
  D.~Israel, C.~Kounnas, D.~Orlando and P.~M.~Petropoulos,
  ``Electric / magnetic deformations of S**3 and AdS(3), and geometric
  cosets,''
  Fortsch.\ Phys.\  {\bf 53}, 73 (2005)
  [arXiv:hep-th/0405213].
  %%CITATION = HEP-TH 0405213;%%
}

%\BanksYZ
\lref\BanksYZ{
  T.~Banks and L.~J.~Dixon,
  ``Constraints On String Vacua With Space-Time Supersymmetry,''
  Nucl.\ Phys.\ B {\bf 307}, 93 (1988).
  %%CITATION = NUPHA,B307,93;%%
}

%\DineVF
\lref\DineVF{
  M.~Dine and N.~Seiberg,
  ``Microscopic Knowledge From Macroscopic Physics In String Theory,''
  Nucl.\ Phys.\ B {\bf 301}, 357 (1988).
  %%CITATION = NUPHA,B301,357;%%
}

%\KutasovUA
\lref\KutasovUA{
  D.~Kutasov and N.~Seiberg,
  ``Noncritical Superstrings,''
  Phys.\ Lett.\ B {\bf 251}, 67 (1990).
  %%CITATION = PHLTA,B251,67;%%
}

%\ItzhakiTU
\lref\ItzhakiTU{
  N.~Itzhaki, D.~Kutasov and N.~Seiberg,
  ``I-brane dynamics,''
  JHEP {\bf 0601}, 119 (2006)
  [arXiv:hep-th/0508025].
  %%CITATION = HEP-TH 0508025;%%
}

%\ItzhakiZR
\lref\ItzhakiZR{
  N.~Itzhaki, D.~Kutasov and N.~Seiberg,
  ``Non-supersymmetric deformations of non-critical superstrings,''
  JHEP {\bf 0512}, 035 (2005)
  [arXiv:hep-th/0510087].
  %%CITATION = HEP-TH 0510087;%%
}

%\HassanGI
\lref\HassanGI{
  S.~F.~Hassan and A.~Sen,
  ``Marginal deformations of WZNW and coset models from O(d,d)
  transformation,''
  Nucl.\ Phys.\ B {\bf 405}, 143 (1993)
  [arXiv:hep-th/9210121].
  %%CITATION = HEP-TH 9210121;%%
}

%\GiveonPH
\lref\GiveonPH{
  A.~Giveon and E.~Kiritsis,
  ``Axial vector duality as a gauge symmetry and topology change in
string
  theory,''
  Nucl.\ Phys.\ B {\bf 411}, 487 (1994)
  [arXiv:hep-th/9303016].
  %%CITATION = HEP-TH 9303016;%%
}

%\KiritsisIU
\lref\KiritsisIU{
  E.~Kiritsis and C.~Kounnas,
  ``Infrared behavior of closed superstrings in strong magnetic and
  gravitational fields,''
  Nucl.\ Phys.\ B {\bf 456}, 699 (1995)
  [arXiv:hep-th/9508078].
  %%CITATION = HEP-TH 9508078;%%
}

%\SuyamaXK
\lref\SuyamaXK{
  T.~Suyama,
  ``Deformation of CHS model,''
  Nucl.\ Phys.\ B {\bf 641}, 341 (2002)
  [arXiv:hep-th/0206171].
  %%CITATION = HEP-TH 0206171;%%
}

%\ForsteKM
\lref\ForsteKM{
  S.~Forste and D.~Roggenkamp,
  ``Current current deformations of conformal field theories, and WZW
  models,''
  JHEP {\bf 0305}, 071 (2003)
  [arXiv:hep-th/0304234].
  %%CITATION = HEP-TH 0304234;%%
}

%\OhmoriAM
\lref\OhmoriAM{
  K.~Ohmori,
  ``A review on tachyon condensation in open string field theories,''
  arXiv:hep-th/0102085.
  %%CITATION = HEP-TH 0102085;%%
}

%\TaylorGN
\lref\TaylorGN{
  W.~Taylor and B.~Zwiebach,
  ``D-branes, tachyons, and string field theory,''
  arXiv:hep-th/0311017.
  %%CITATION = HEP-TH 0311017;%%
}

%\MartinecTZ
\lref\MartinecTZ{
  E.~J.~Martinec,
  ``Defects, decay, and dissipated states,''
  arXiv:hep-th/0210231.
  %%CITATION = HEP-TH 0210231;%%
}

%\HeadrickHZ
\lref\HeadrickHZ{
  M.~Headrick, S.~Minwalla and T.~Takayanagi,
  ``Closed string tachyon condensation: An overview,''
  Class.\ Quant.\ Grav.\  {\bf 21}, S1539 (2004)
  [arXiv:hep-th/0405064].
  %%CITATION = HEP-TH 0405064;%%
}

%\SeibergBX
\lref\SeibergBX{
  N.~Seiberg,
  ``Observations on the moduli space of two dimensional string
theory,''
  JHEP {\bf 0503}, 010 (2005)
  [arXiv:hep-th/0502156].
  %%CITATION = HEP-TH 0502156;%%
}

%\FotopoulosCN
\lref\FotopoulosCN{
  A.~Fotopoulos, V.~Niarchos and N.~Prezas,
  ``D-branes and SQCD in non-critical superstring theory,''
  JHEP {\bf 0510}, 081 (2005)
  [arXiv:hep-th/0504010].
  %%CITATION = HEP-TH 0504010;%%
}

%\IsraelFN
\lref\IsraelFN{
  D.~Israel, A.~Pakman and J.~Troost,
  ``D-branes in little string theory,''
  Nucl.\ Phys.\ B {\bf 722}, 3 (2005)
  [arXiv:hep-th/0502073].
  %%CITATION = HEP-TH 0502073;%%
}

%\AshokPY
\lref\AshokPY{
  S.~K.~Ashok, S.~Murthy and J.~Troost,
  ``D-branes in non-critical superstrings and minimal super Yang-Mills
in
  various dimensions,''
  arXiv:hep-th/0504079.
  %%CITATION = HEP-TH 0504079;%%
}

%\MurthyEG
\lref\MurthyEG{
  S.~Murthy,
  ``Non-critical heterotic superstrings in various dimensions,''
  arXiv:hep-th/0603121.
  %%CITATION = HEP-TH 0603121;%%
}

%\AharonyUB
\lref\AharonyUB{
  O.~Aharony, M.~Berkooz, D.~Kutasov and N.~Seiberg,
  ``Linear dilatons, NS5-branes and holography,''
  JHEP {\bf 9810}, 004 (1998)
  [arXiv:hep-th/9808149].
  %%CITATION = HEP-TH 9808149;%%
}

%\KazamaQP
\lref\KazamaQP{
  Y.~Kazama and H.~Suzuki,
  ``New N=2 Superconformal Field Theories And Superstring
Compactification,''
  Nucl.\ Phys.\ B {\bf 321}, 232 (1989).
  %%CITATION = NUPHA,B321,232;%%
}

%\HoriAX
\lref\HoriAX{
  K.~Hori and A.~Kapustin,
  ``Duality of the fermionic 2d black hole and N = 2 Liouville theory
as
  mirror symmetry,''
  JHEP {\bf 0108}, 045 (2001)
  [arXiv:hep-th/0104202].
  %%CITATION = HEP-TH 0104202;%%
}

%\BakasBA
\lref\BakasBA{
  I.~Bakas,
  ``Space-time interpretation of s duality and supersymmetry violations
of t
  duality,''
  Phys.\ Lett.\ B {\bf 343}, 103 (1995)
  [arXiv:hep-th/9410104].
  %%CITATION = HEP-TH 9410104;%%
}

%\BergshoeffCB
\lref\BergshoeffCB{
  E.~Bergshoeff, R.~Kallosh and T.~Ortin,
  ``Duality versus supersymmetry and compactification,''
  Phys.\ Rev.\ D {\bf 51}, 3009 (1995)
  [arXiv:hep-th/9410230].
  %%CITATION = HEP-TH 9410230;%%
}

%\BakasHC
\lref\BakasHC{
  I.~Bakas and K.~Sfetsos,
  ``T duality and world sheet supersymmetry,''
  Phys.\ Lett.\ B {\bf 349}, 448 (1995)
  [arXiv:hep-th/9502065].
  %%CITATION = HEP-TH 9502065;%%
}

%\IsraelIR
\lref\IsraelIR{
  D.~Israel, C.~Kounnas, A.~Pakman and J.~Troost,
  ``The partition function of the supersymmetric two-dimensional black
hole
  and little string theory,''
  JHEP {\bf 0406}, 033 (2004)
  [arXiv:hep-th/0403237].
  %%CITATION = HEP-TH 0403237;%%
}

%\AntoniadisSW
\lref\AntoniadisSW{
  I.~Antoniadis, S.~Dimopoulos and A.~Giveon,
  ``Little string theory at a TeV,''
  JHEP {\bf 0105}, 055 (2001)
  [arXiv:hep-th/0103033].
  %%CITATION = HEP-TH 0103033;%%
}

%\ChaudhuriQB
\lref\ChaudhuriQB{
  S.~Chaudhuri and J.~A.~Schwartz,
  ``A Criterion For Integrably Marginal Operators,''
  Phys.\ Lett.\ B {\bf 219}, 291 (1989).
  %%CITATION = PHLTA,B219,291;%%
}

%\DijkgraafVP
\lref\DijkgraafVP{
  R.~Dijkgraaf, E.~P.~Verlinde and H.~L.~Verlinde,
  ``C = 1 Conformal Field Theories On Riemann Surfaces,''
  Commun.\ Math.\ Phys.\  {\bf 115}, 649 (1988).
  %%CITATION = CMPHA,115,649;%%
}

%\GinspargEB
\lref\GinspargEB{
  P.~H.~Ginsparg,
  ``Curiosities At C = 1,''
  Nucl.\ Phys.\ B {\bf 295}, 153 (1988).
  %%CITATION = NUPHA,B295,153;%%
}

%\DixonAC
\lref\DixonAC{
  L.~J.~Dixon, P.~H.~Ginsparg and J.~A.~Harvey,
  ``(Central Charge C) = 1 Superconformal Field Theory,''
  Nucl.\ Phys.\ B {\bf 306}, 470 (1988).
  %%CITATION = NUPHA,B306,470;%%
}

%\LinNH
\lref\LinNH{
  H.~Lin and J.~Maldacena,
  ``Fivebranes from gauge theory,''
  arXiv:hep-th/0509235.
  %%CITATION = HEP-TH 0509235;%%
}

%\IsraelZP
\lref\IsraelZP{
  D.~Israel,
  ``Non-critical string duals of N = 1 quiver theories,''
  arXiv:hep-th/0512166.
  %%CITATION = HEP-TH 0512166;%%
}

%\GiveonFU
\lref\GiveonFU{
  A.~Giveon, M.~Porrati and E.~Rabinovici,
  ``Target space duality in string theory,''
  Phys.\ Rept.\  {\bf 244}, 77 (1994)
  [arXiv:hep-th/9401139].
  %%CITATION = HEP-TH 9401139;%%
}

%\MurthyES
\lref\MurthyES{
  S.~Murthy,
  ``Notes on non-critical superstrings in various dimensions,''
  JHEP {\bf 0311}, 056 (2003)
  [arXiv:hep-th/0305197].
  %%CITATION = HEP-TH 0305197;%%
}

%\SfetsosXD
\lref\SfetsosXD{
  K.~Sfetsos,
  ``Branes for Higgs phases and exact conformal field theories,''
  JHEP {\bf 9901}, 015 (1999)
  [arXiv:hep-th/9811167].
  %%CITATION = HEP-TH 9811167;%%
}

%\KutasovPV
\lref\KutasovPV{
  D.~Kutasov,
  ``Some properties of (non)critical strings,''
  arXiv:hep-th/9110041.
  %%CITATION = HEP-TH 9110041;%%
}

%\CallanAT
\lref\CallanAT{
  C.~G.~.~Callan, J.~A.~Harvey and A.~Strominger,
  ``Supersymmetric string solitons,''
  arXiv:hep-th/9112030.
  %%CITATION = HEP-TH 9112030;%%
}

%\HarmarkDT
\lref\HarmarkDT{
  T.~Harmark and N.~A.~Obers,
  ``Thermodynamics of the near-extremal NS5-brane,''
  Nucl.\ Phys.\ B {\bf 742}, 41 (2006)
  [arXiv:hep-th/0510098].
  %%CITATION = HEP-TH 0510098;%%
}

%%%%%%%%%%%%%%%%%%%%%%
%                      Title Page                                 %
%%%%%%%%%%%%%%%%%%%%%%
\Title{
\vbox{%\hbox{EFI-03-41}
      \hbox{hep-th/0605192}}
}
{\vbox{
\centerline{Stable Non-Supersymmetric Vacua in the}
\vskip0.4cm
\centerline{Moduli Space of Non-Critical Superstrings}}}

\vskip .1in

\centerline{Troels Harmark\footnote{$^{a}$}{harmark@nbi.dk},
Vasilis Niarchos\footnote{$^{b}$}{niarchos@nbi.dk} and
Niels A.\ Obers\footnote{$^{c}$}{obers@nbi.dk}}

\vskip .2in

%\vskip 2cm
\centerline{{\it The Niels Bohr Institute}}
\centerline{\it Blegdamsvej 17, 2100 Copenhagen \O, Denmark}

\vskip 1.2cm
\noindent

%%abstract

\noindent
We study a set of asymmetric deformations
of non-critical superstring theories in various dimensions. The
deformations arise as K\"ahler and complex structure deformations
of an orthogonal two-torus comprising of a parallel and a transverse direction
in the near-horizon geometry of NS5-branes. The resulting theories
have the following intriguing features: Spacetime
supersymmetry is broken in a continuous fashion and the masses of the lightest modes
are lifted. In particular, no bulk or localized tachyons are generated in the
non-supersymmetric vacua. We discuss the effects of these
deformations in the context of the holographic duality between non-critical
superstrings and Little String Theories and find solutions of rotating
fivebranes in supergravity. We also comment on the generation
of a one-loop cosmological constant and determine the effects of the
one-loop backreaction to leading order.

\Date{May, 2006}

\vfill
\vfill

\listtoc
\writetoc

\newsec{Introduction}

String theory with broken supersymmetry exhibits a variety of interesting features.
At tree-level, the theory can develop perturbative instabilities, which are signaled by the
presence of tachyons, $i.e.$ particles with negative mass squared,
in the perturbative spectrum. The condensation of these modes is a
time-dependent process driving the theory towards a more stable
vacuum, where some amount of supersymmetry is usually restored.
In recent years, much progress has been achieved in understanding such
processes in open string theory (for reviews see
\refs{\OhmoriAM,\TaylorGN}). A corresponding analysis of bulk tachyon
dynamics is significantly more involved and a general treatment is
still lacking (see, however, \refs{\MartinecTZ,\HeadrickHZ} and references therein).
At  higher loops, an effective potential is generated, which typically
lifts some of the flat directions of the theory and produces a non-vanishing
cosmological constant. Features like broken supersymmetry, time dependence
and a non-vanishing cosmological constant are expected to be standard properties
of any fundamental theory of the real world. Hence, it is imperative to
understand these and other dynamical aspects of supersymmetry breaking in string theory.

In standard compactifications of string theory a successful
resolution of the hierarchy problem requires that we break
supersymmetry at a scale $m_{SUSY}$ far below the string scale $m_s$.
Usually, when we break supersymmetry at tree-level the supersymmetry
breaking scale is closely tied to the compactification scale $m_c$.
The Scherk-Schwarz mechanism is a typical example of this property.
At weak $g_s$ coupling the compactification scale is itself comparable
to the string scale which makes it difficult to generate a large
separation of scales between $m_{SUSY}$ and $m_s$
\refs{\KiritsisMC,\PolchinskiRR}.
Supersymmetry breaking at higher loops in perturbation theory or
by non-perturbative effects, such as gaugino condensation,
instantons $etc.$ can avoid this problem and is
very attractive for phenomenological applications, but takes us into the
realm of strong coupling dynamics and will not be discussed here.

The simplest, and most obvious way, to obtain a small
$m_{SUSY}/m_s$ ratio and a correspondingly small one-loop cosmological
constant, is to start with a supersymmetric vacuum and continuously turn on
a modulus that breaks the supersymmetry. Thus, vacua in the
neighborhood of the supersymmetric point would exhibit an arbitrarily small
breaking of supersymmetry. There are, however, general arguments
in standard string theory compactifications \refs{\BanksYZ,\DineVF,\PolchinskiRR}
that exclude this possibility. The Scherk-Schwarz
supersymmetry breaking mechanism is again a nice example of the
generic situation. Supersymmetry is explicitly broken at any finite
compactification radius and is only restored when the radius becomes
infinite, $i.e.$ at infinite distance in the moduli space.

An interesting loophole to the general arguments of the previous
paragraph can be found in string theories that live on asymptotically linear
dilaton backgrounds. These are typically backgrounds of the general form
\eqn\aaa{
\IR^{d-1,1}\times \IR_{\phi}\times \NN
~,}
where $\IR^{d-1,1}$ is the $d$-dimensional Minkowski spacetime,
$\IR_{\phi}$ is a linear dilaton direction labeled by $\phi$ and $\NN$ is a compact
space. The string coupling $g_s$ vanishes at the asymptotic boundary
$\phi \rightarrow \infty$ and grows exponentially as we move towards
smaller and smaller $\phi$. With an even number of
Minkowski spacetime dimensions $d$,\foot{$d=2n$,
with $n=0,1,\cdots,4$. $d=8$ is the ten-dimensional critical string.}
a special class of solutions that preserves at least
$2^{\frac{d}{2}+1}$
spacetime supersymmetries has been considered in \KutasovUA.
In these solutions \aaa\ takes the form
\eqn\aab{
\IR^{d-1,1}\times \IR_{\phi}\times (S^1 \times \MM)/\Gamma
~,}
where $\MM$ is the target space of a two-dimensional CFT with
$\NN=(2,2)$ worldsheet supersymmetry and $\Gamma$
is a discrete group associated with the chiral
GSO projection. String theory on \aab\ appears naturally in the near horizon
limit of NS5-branes in type II string theory and defines the holographic dual
of Little String Theory (LST) \refs{\AharonyKS,\KutasovUF}.

It is important to stress that the radius of the $S^1$ that appears
in \aab\ is not arbitrary, but is fixed by the GSO projection
in terms of the linear dilaton slope. Only at this special radius
is string theory on \aab\ spacetime supersymmetric.
Thus, a natural way to break the spacetime supersymmetry
continuously (and evade the general arguments of
\refs{\BanksYZ,\DineVF,\PolchinskiRR}) is to turn on the modulus that changes
the radius of the $S^1$ \KutasovUA. The resulting moduli space of vacua has been
studied recently in \ItzhakiZR, following earlier work on the $d=0$ case in \SeibergBX.

The stability properties of the vacua obtained by the above
non-supersymmetric deformations was one of the main issues
analyzed in \ItzhakiZR. In spacetimes of the form \aaa, there are
in general two kinds of instabilities that can appear when we break
the spacetime supersymmetry. One kind corresponds to delta function
normalizable states propagating in the bulk of the linear dilaton throat $\IR_{\phi}$.
Henceforth, we will refer to such instabilities as bulk tachyons. The other kind
is characterized by normalizable states localized deep inside the
strongly coupled region of the throat. We will refer to such instabilities as
localized tachyons.

The analysis of \ItzhakiZR\ revealed that the stability properties of
string theory in the moduli space of \aab\ depend crucially on the
flat spacetime dimension $d$ and the compact manifold $\MM$.
In the vicinity of the supersymmetric point and for $d=2$
the theory is free of both bulk and localized tachyons.
The same is true also for $d=3$, an interesting odd $d$
case that describes the near horizon region of a system
of intersecting NS5-branes \ItzhakiTU. For $d=4$ and $\MM=0$,
a case that describes the decoupling limit of the
conifold singularity, the theory exhibits both bulk and localized tachyons
arbitrarily close to the supersymmetric point. Finally, for $d=6$ and $\MM=0$ or
$\MM=\frac{SU(2)}{U(1)}$, the theory exhibits a localized tachyon, but
no bulk tachyons in a finite region in the moduli space around the supersymmetric point.

In this paper, we show that there is a larger moduli space of
non-supersymmetric vacua around the supersymmetric theory on \aab.
In order to explain the general situation, let us consider the
following prototypical example. For $d=6$, $\MM=0$ and
$R_Y=Q=1/\sqrt{2}$ ($R_Y$ is the radius of the $S^1$ labeled
here by $Y$ and $Q$ is the linear dilaton slope)
the background \aab\ preserves sixteen supersymmetries and is relevant
for the near-horizon dynamics of two parallel NS5-branes. Compactifying one
of the worldvolume directions (let us call it $X$) preserves the same amount
of supersymmetry and gives rise to the non-critical background
\eqn\aac{
\IR^{4,1}\times \IR_{\phi}\times S^1_X \times S^1_Y
~.}
{} From the five-dimensional point of view we can write the spacetime
supercharges as
\eqn\aada{
Q^1_{\alpha}=\oint \frac{dz}{2\pi i}~
e^{-\frac{\varphi}{2}+\frac{i}{2}(H_2+H_3-\sqrt 2 Y)
+i\frac{\alpha}{2}(H_0+H_1)},
Q^1_{\dot{\alpha}}=\oint \frac{dz}{2\pi i}~
e^{-\frac{\varphi}{2}-\frac{i}{2}(H_2+H_3-\sqrt 2 Y)
+i\frac{\dot{\alpha}}{2}(H_0-H_1)}
,}
\eqn\aadb{
Q^2_{\alpha}=\oint \frac{dz}{2\pi i}~
e^{-\frac{\varphi}{2}+\frac{i}{2}(-H_2+H_3-\sqrt 2 Y)
+i\frac{\alpha}{2}(-H_0+H_1)},
Q^2_{\dot{\alpha}}=\oint \frac{dz}{2\pi i}~
e^{-\frac{\varphi}{2}+\frac{i}{2}(H_2-H_3+\sqrt 2 Y)
-i\frac{\dot{\alpha}}{2}(H_0+H_1)}
,}
where $\alpha=\pm$, $\dot{\alpha}=\pm$ and $H_2, H_3$ are respectively
the bosons bosonizing the worldsheet fermions $(\psi^X,\psi^{x^4})$
and $(\psi^Y,\psi^{\phi})$. More details on our conventions and the
spacetime supersymmetry algebra can be found in the next section and
appendix A. An additional equal number of spacetime supercharges
$\bar Q^i_{\alpha}, \bar Q^i_{\dot{\alpha}}$ ($i=1,2$) arises from the right-moving
sector.

The two-torus $S^1_X\times S^1_Y$ of the asymptotic geometry \aac\
exhibits four independent exactly marginal deformations (the usual two complex
and two K\"ahler structure deformations of the two-torus). The first is
given by the worldsheet interaction $\int d^2z ~ \d X\bar \d X$ and corresponds to
the modulus that changes the compactification radius $R_X$. This modulus commutes
with the spacetime supercharges \aada, \aadb\ and, as expected, does not
break any spacetime supersymmetry. The second deformation is given by the
worldsheet interaction $\int d^2z ~ \d Y\bar \d Y$ and corresponds to the modulus
that changes the $S^1_Y$ radius $R_Y$. This is precisely the deformation analyzed in
\ItzhakiZR. As we see explicitly here, this modulus does not commute with any of
the supercharges \aada, \aadb\ and breaks the spacetime supersymmetry
completely. The remaining two-parameter family of deformations is given by the
(asymmetric) worldsheet interaction
\eqn\aae{
\delta \SS_{(\lambda_+,\lambda_-)} \propto
\int d^2z~ \Big( \lambda_+ \OO_+(z,\bar z)+\lambda_- \OO_-(z,\bar z)\Big)
~,}
where
\eqn\aaf{
\OO_{\pm}=\d X \bar \d Y \pm \d Y \bar \d X
~.}
This is a modulus that turns on a constant off-diagonal component of
the metric and/or a constant B-field. As we explain in the main text, this perturbation can be
viewed as giving an expectation value to the corresponding bulk gauge fields, {\it i.e.}\
turning on an appropriate Wilson line.

For general values of the deformation parameters $\lambda_{\pm}$ with
$|\lambda_+|\neq |\lambda_-|$ the worldsheet interaction
$\delta \SS_{(\lambda_+,\lambda_-)}$ does not commute with any of
the spacetime supercharges \aada, \aadb\ and like
$\int d^2z ~ \d Y\bar \d Y$ it breaks the spacetime supersymmetry
completely. On the special one-dimensional submanifold $|\lambda_+|=|\lambda_-|$
the deformations break only half of the supersymmetry.
For concreteness, we denote these special deformations as
\eqn\aaga{
\delta \SS_{\lambda} \propto \frac{\lambda}{2}\int d^2 z~ \Big(\OO_+
+\OO_-\Big)=
\lambda\int d^2 z~ \d X \bar \d Y
~,}
\eqn\aagb{
\delta \SS_{\bar \lambda} \propto \frac{\bar \lambda}{2}\int d^2 z~
\Big(\OO_+ -\OO_-\Big)=
\bar \lambda\int d^2 z~ \d Y \bar \d X
~.}
$\delta \SS_{\lambda}$ preserves the eight supercharges $Q^i_{\alpha}$,
$Q^i_{\dot \alpha}$ arising from the left-moving sector of the worldsheet,
and $\delta \SS_{\bar \lambda}$ preserves the other eight supercharges
$\bar Q^i_{\alpha}, \bar Q^i_{\dot{\alpha}}$ arising from the
right-moving sector.

It is worth mentioning that a set of asymmetric deformations,
very similar to our $\delta \SS_{\lambda}$, $\delta \SS_{\bar \lambda}$
above, has been considered also in the past in the context of heterotic
$SU(2)$ and $SL(2)$ WZW models \refs{\IsraelVV,\IsraelCD}. There it was argued that
an appropriate asymmetric deformation gives rise to a line of exact
conformal field theories interpolating between the $SU(2)$ (or $SL(2)$)
WZW model and the CFT that describes string propagation on the geometric
coset $S^2$ (or $AdS_2$).

The main purpose of this paper is to analyze the effect of the general
asymmetric deformation \aae\ on type II string theory on spacetimes
of the form \aab. For concreteness, we will focus on the $d=6$ case
with $\MM=0$ or $\MM=SU(2)/U(1)$ and will compactify
one of the flat spatial $\IR^{d-1}$ directions, which we will call in
general $X$. We should emphasize, however,
that similar results can be obtained for any other $d>0$ and $\MM$
and for asymmetric deformations involving any of the flat
$\IR^{d-1,1}$ directions. Some of the possible extensions will be
discussed briefly in section 6.

Special emphasis will be given to the stability properties of the
moduli space that arises by turning on the asymmetric deformations.
One of the striking results of our analysis is that, contrary to the
symmetric $\int d^2 z ~\d Y\bar \d Y$ deformation, the asymmetric deformations
always give non-negative contributions to the masses of the lightest modes and
bulk or localized tachyons do not appear. This surprising feature
is independent of the dimension $d$ and allows
us to construct tree-level stable non-supersymmetric string theories with
asymptotic linear dilaton directions. A detailed analysis of the effect
of the asymmetric deformations on the spectrum of the theory appears in
section 2.

Another interesting aspect of our analysis has to do with holography.
As we mentioned earlier, string theory on \aab\ defines the
holographic dual of a $d$-dimensional non-gravitational (and non-local) theory
known as Little String Theory. Many aspects of holography in
this setting have been discussed in a series of papers
\refs{\AharonyUB\GiveonPX-\AharonyXN}.  It is known, in particular,
how to map in the type IIB case a class of spacetime chiral primary states in the
non-critical string description to low-energy gauge theory operators in
the S-dual D5-brane gauge theory. It is interesting to identify the leading
order effects of the asymmetric deformation \aae\ of the bulk theory in the dual D5-brane gauge
theory. Using the general holographic prescription, the deformation amounts
in the gauge theory to adding appropriate R-symmetry currents. This is explicitly
verified in section 3 with a D5-brane probe analysis in the S-dual 
of the deformed non-critical string background.
Furthermore, we will see that the results of the DBI
analysis are in good agreement with the leading order
spectrum deformations in section 2. A similar analysis for the symmetric
deformation $\int d^2 z ~\d Y\bar \d Y$ was performed in \ItzhakiZR.
In addition, we find in general that the lightest modes in the bulk
receive their first mass squared contribution at second order
in the deformation parameter. On the D5-brane side, this implies
a quadratic potential interaction, which is a non-chiral operator in a
non-supersymmetric theory. The D5-brane probe analysis also reproduces this
term.

In section 4 we explore the physics of
the Higgs branch of the dual LST. At the supersymmetric point ($i.e$
before the asymmetric deformation), we can regularize the strong coupling
singularity of the backgrounds \aab, by turning on the appropriate $\NN=2$
Liouville interaction on the worldsheet theory. In the language of NS5-branes this deformation takes
us into the Higgs branch, where the NS5-branes are separated in an
appropriate double scaling limit \GiveonPX. In the presence of the asymmetric
deformation, the $\NN=2$ Liouville interaction is irrelevant (massive)
on the worldsheet and the true exactly marginal Liouville interaction
is time-dependent. In section 4, we find exact supergravity backgrounds
in the large $k$ limit, which describe the non-trivial rotation of $k$ parallel
NS5-branes in the presence of the asymmetric deformation. We deduce the
precise form of these rotating solutions with a sequence of
T-dualities and boosts. Similar solutions for $\int d^2 z ~\d Y\bar \d Y$
appeared in \ItzhakiZR.

In section 5, we discuss some of the features of the one-loop
backreaction problem for the non-supersymmetric deformations and the
generation of a one-loop cosmological constant. We explicitly find for the deformed
CHS background the leading order correction coming from the one-loop backreaction.
In section 6, we discuss some immediate extensions of our work and give
an overview of the larger moduli space  of string theory on \aab. We conclude in section 7
with a brief summary and some interesting prospects. Two appendices are included
summarizing our notation and some facts that are useful in the main text.

\newsec{Deformations of the spectrum}

In this section we define the asymmetric deformations of interest more
generally and analyze their effect on
the spectrum of the theory for a special class of backgrounds of the
form \aab\ with $d=6$ and $\MM=SU(2)/U(1)$. Similar results can be obtained
also in other dimensions and with different compact spaces $\MM$.
Possible extensions will be discussed in section 6.

\subsec{Setting the stage}

A specific example of the general form \aaa\ is the CHS background \CallanAT
\eqn\baa{ \IR^{5,1}\times \IR_{\phi} \times SU(2)_k
~,}
which appears in the near horizon geometry of $k$ parallel NS5-branes.
The linear dilaton slope depends on the number of NS5-branes via
the relation\foot{We use the convention
$\alpha'=1$ in this paper. Our notation is summarized in appendix A.}
\eqn\bab{
Q=1/\sqrt{k}
~.}
For what follows, it will be convenient
to single out one of the flat directions labeled by $X$ and
compactify it with an arbitrary radius $R_X$. We will denote the
remaining flat directions as $x^{\mu}$ with $\mu=0,1,...,4$. In
addition, the worldsheet theory on \baa\ comprises of a set of
free real fermions $\psi^{\mu}$, $\psi^X$, $\psi^{\phi}$ and the
$\NN=1$ supersymmetric WZW model $SU(2)_k$. The latter can be
written as a direct sum of the bosonic $SU(2)_{k-2}$ WZW
model and three free fermions $\chi^3,\chi^{\pm}$.

The CHS background \baa\ can also be recast as
\eqn\bac{
\IR^{5,1}\times \IR_{\phi} \times \bigg(S^1_k \times
\frac{SU(2)_k}{U(1)}\bigg)/\IZ_k
}
by using the well-known decomposition of the $SU(2)_k$
\eqn\bad{
SU(2)_k=\bigg(S^1_k\times \frac{SU(2)_k}{U(1)}\bigg)/\IZ_k
}
in terms of the $\NN=2$ minimal model $SU(2)_k/U(1)$. Hence, \bac\ is a
special case of the general solution \aab\ with $\MM_k=SU(2)_k/U(1)$
and $\Gamma=\IZ_k$. Notice that for $k=2$ the minimal model $\MM_2$ is
empty and \bac\ becomes simply
\eqn\bae{
\IR^{5,1}\times \IR_{\phi}\times S^1
~.}

At the supersymmetric point, the radius of the $S^1$ in \bac\ is
fixed by the GSO projection to be $Q$ or $1/Q$ (the two are related
by T-duality). By changing this radius we break the spacetime
supersymmetry and move away from the supersymmetric point
in the moduli space. The properties of this deformation were analyzed
extensively in \ItzhakiZR. The worldsheet modulus that is responsible
for this effect can be written as
\eqn\baf{
\delta \SS=\lambda \int d^2 z ~ \d Y \bar \d Y
~,}
where $Y$ is the boson that labels the $S^1$ in \bac. In the $SU(2)$
formulation \baa\ we can think of $Y$ as the boson that bosonizes
the total $K_3^{(tot)}$ current of the $SU(2)_k$ WZW model, $i.e.$
\eqn\bag{
K_3^{(tot)}=K_3+\chi^+\chi^-=i\sqrt{k}\d Y
~.}
In this expression $K_3$ is the Cartan current of the $bosonic$
$SU(2)_{k-2}$ current algebra. Thus, in the CHS geometry \baa\ we can
write the deformation \baf\ as
\eqn\bai{
\delta \SS=-\lambda \int d^2 z~ G_{-1/2} \bar G_{-1/2} ~
\chi^3\bar\chi^3=
-Q^2 \lambda \int d^2 z~ K_3^{(tot)}(z)\bar K_3^{(tot)}(\bar z)=
\lambda \int d^2 z ~\d Y\bar \d Y }
where $G_{-1/2}$, $\bar G_{-1/2}$ are the fermionic generators
of the $\NN=1$ worldsheet supersymmetry algebra (see appendix A).
This particular current-current deformation has been discussed in
various papers in the past
\refs{\HassanGI\GiveonPH\KiritsisIU\SuyamaXK-\ForsteKM}.
{} From the spacetime point of view, it is a deformation that
squashes the three-sphere transverse to the fivebranes and since it
does not commute with the spacetime supercharges it also
breaks the spacetime supersymmetry explicitly.

In this paper, we are interested in another set of current-current
deformations, which are left-right asymmetric and couple the parallel and transverse
directions of the fivebranes in an interesting fashion. They have the general
form\foot{Anticipating the discussion of the next section it is convenient to label
the deformation parameters here as $\tilde \lambda_{\pm}$.}
\eqn\baj{\eqalign{
\delta \SS_{(\tilde \lambda_+,\tilde \lambda_-)}&=
-\frac{\tilde \lambda_++\tilde \lambda_-}{4\pi}\int d^2z~  G_{-1/2}
\bar G_{-1/2} ~ \psi^X \bar \chi^3
-\frac{\tilde \lambda_+-\tilde \lambda_-}{4\pi}\int d^2z~ G_{-1/2} \bar
G_{-1/2} ~ \chi^3\bar \psi^X=
\cr
&=\frac{1}{4\pi}\int d^2z \Big(\tilde \lambda_+ \OO_+(z,\bar z) +\tilde
\lambda_- \OO_-(z,\bar z)\Big)
~,}}
with
\eqn\bak{
\OO_{\pm}=\d X\bar \d Y\pm \d Y\bar \d X
~.}
The $\frac{1}{4\pi}$ factor has been inserted for later convenience and
the bosons $X,Y$ satisfy the periodicity conditions $X \sim X + 2\pi R_X$, $Y \sim Y + 2 \pi R_Y$.
As we mentioned in the introduction, for generic values of the
deformation parameters $\tilde \lambda_{\pm}$  the modulus \baj\ breaks the
spacetime supersymmetry completely. The deformation preserves half of the spacetime
supersymmetry when $|\tilde \lambda_+|=|\tilde \lambda_-|$
(see appendix A for the explicit form of the spacetime supercharges).
Our aim in this section is to determine the effect of the
deformations \baj, \bak\ on the spectrum of the supersymmetric theory.

The vertex operators that appear in \baj\ express the $x$-components
$({\cal{A}}_{L})_x^3$ and $({\cal{A}}_{R})_x^3$ of the $SU(2)_{L,R}$ bulk gauge fields
$({\cal{A}}_{L,R})_\mu^a$ ($\mu = 0 \ldots 5,a=3,\pm $). As a consequence,
 the asymmetric deformation \baj\ can also be written as
\eqn\defg{
\delta \SS_{(\tilde \lambda_+,\tilde \lambda_-)} \propto \int d^2 z \left(
\tilde \lambda_- [({\cal{A}}_{R})_x^3 - ({\cal{A}}_{L})_x^3 ]
+ \tilde \lambda_+ [({\cal{A}}_{R})_x^3 + ({\cal{A}}_{L})_x^3 ] \right) ~ 
}
and hence can be viewed as giving an expectation value to these gauge
 field components.

\subsec{O(2,2) deformations of the spectrum}

Let us begin by recalling the main features of the spectrum at the
supersymmetric point. The $k=2$ case \bae\ is simpler and we
review it first. In the linear dilaton geometry \bae\ the general vertex
operator (without any string oscillators) takes the form
\eqn\bba{
\VV=e^{-(1-\frac{\alpha}{2})\varphi-(1-\frac{\bar \alpha}{2})\bar
\varphi}
~ e^{i \sum_{i=0}^3 (s_i H_i+\bar s_i\bar H_i)}
~ e^{i(pX+\bar p \bar X)+i(qY+\bar q \bar Y)}
~ e^{ip_{\mu} x^{\mu}} e^{\beta \phi}
,}
where $\alpha=0$ stands for the NS sector and $\alpha=1$ for the
R sector. $\varphi$ is the standard $(\beta,\gamma)$ superghost
of the fermionic string and $H_i$ ($i=0,1,2,3$) are four bosons
bosonizing the eight free worldsheet fermions of the theory (more
details can be found in appendix A).

Following the notation of \PolchinskiRR\
we denote the general closed string sector of the theory by
\eqn\bbb{
(\alpha, F,\bar \alpha, \bar F)
~,}
where $\alpha$, $\bar \alpha$ are the labels appearing in \bba\
and $F, \bar F$ are the left- and right-moving worldsheet fermion
numbers modulo two. In the $d=6$ case that we study here $F=0,1$ in the
NS sector and $\pm \frac{1}{2}$ in the Ramond sector.

The allowed quantum numbers in the supersymmetric theory
are determined by the chiral GSO projection ($i.e.$ the requirement of
locality of the generic vertex operator \bba\ with the spacetime supercharges),
the mutual locality conditions and the physical state conditions.
In the present case, the mutual locality condition reads
\eqn\bbc{
F_1 \alpha_2-F_2 \alpha_1-\bar F_1 \bar \alpha_2+
\bar F_2 \bar \alpha_1-\frac{1}{2}(\alpha_1\alpha_2-\bar \alpha_1\bar
\alpha_2)
+2(n_1 w_2+n_2 w_1)\in 2 \IZ
~,}
where $n,w$ are the momentum and winding quantum numbers
in the $Y$ direction, $i.e.$ in \bba
\eqn\bbd{
q=\frac{n}{R_Y}+ wR_Y ~, ~ ~
\bar q=\frac{n}{R_Y}- wR_Y
~.}
At the supersymmetric point $R_Y=Q=1/\sqrt{2}$. The allowed sectors
and states can be found in \refs{\KutasovPV,\MurthyES,\ItzhakiZR}.

For $k>2$ the analog of the general vertex operator \bba\ is
\eqn\bbe{
\VV=e^{-(1-\frac{\alpha}{2})\varphi-(1-\frac{\bar \alpha}{2})\bar
\varphi}
~ e^{i \sum_{i=0}^3 (s_i H_i+\bar s_i\bar H_i)}
~ e^{i(pX+\bar p \bar X)+i(qY+\bar q \bar Y)}
~ e^{ip_{\mu} x^{\mu}} e^{\beta \phi}
~ \Phi_{j+1,m,\bar m}
,}
where $\Phi_{j+1,m,\bar m}$ is a primary vertex operator of the
$\NN=2$ supersymmetric  minimal model $\MM_k=SU(2)_k/U(1)$.
The minimal model quantum numbers $(j,m,\bar m)$ can take the values
$j=0,\frac{1}{2},...,\frac{k}{2}-1$, $m,\bar m=-j-1,-j,-j+1,...,j+1$.
The latter are intimately tied to the momentum and
winding quantum numbers $n, w$ along $Y$ through the chiral
GSO projection. $n$ and $w$ are still given by
\bbd, but now $R_Y=Q=1/\sqrt{k}$ and in terms of
$m,\bar m$
\eqn\bbf{
q=2Qm~, ~ ~ \bar q=2Q\bar m
~.}
Notice that the GSO projection allows for a fractional momentum
quantum number $n\in \IZ/k$.

The (five-dimensional) mass of the corresponding spacetime modes
can be determined in the following manner. If we denote by $h$
the (say, left-moving) scaling dimension of the generic vertex operator
$\VV$ in \bbe\
at five-dimensional momentum $p_{\mu}=0$, we can write the mass-shell
condition as
\eqn\bbg{
h-\frac{1}{4}M^2=1 \Leftrightarrow M^2=4(h-1)
~,}
where $M^2=-p_{\mu}p^{\mu}$ is the five-dimensional mass squared.
Notice that we do not have to deal separately with the left- and
right-moving scaling dimensions, because one of the physical state
conditions is $L_0=\bar L_0$. Hence, in order to determine the effect of the
asymmetric deformations \baj\ on the spectrum of the theory, we need to
determine how it affects the scaling dimensions $h$. For example, if we have
initially a massless mode and the asymmetric deformation gives $\delta h<0$,
then a tachyon will appear in the deformed theory. This tachyon
will be a localized tachyon if the quantum number $\beta$ in \bbe\ is
real, or a bulk tachyon if $\beta=-\frac{Q}{2}+is$, $s\in \IR$.

Determining the effect of the asymmetric deformations on the scaling
dimensions $h$ is a straightforward exercise, because the general
deformation \baj\ acts only on the free $(X,Y)$ part the worldsheet
theory. Then, essentially we have to find the scaling dimensions
of the vertex operators
\eqn\bbi{
e^{i(pX+\bar p \bar X)+i(qY+\bar q \bar Y)}
}
under a general $O(2,2)$ deformation of the orthogonal torus
$S^1_X\times S^1_Y$. A general expression for these scaling dimensions
is known.

Consider the general two-torus with metric and $B$-field
\eqn\bbj{
ds^2=G_{xx}dx^2+G_{yy}dy^2+2G_{xy} dxdy
~, ~ ~
B=B_{xy} dx\wedge dy
~.}
where $X= R_X x$, $Y= R_Y y$.
It is convenient to arrange the four real data $G_{xx}, G_{yy},
G_{xy}$, $B_{xy}$ in two complex parameters $\rho$ and $\tau$ in
the following manner
\eqn\bbka{
\tau\equiv \tau_1+i\tau_2=\frac{G_{xy}}{G_{yy}}+i\frac{\sqrt
G}{G_{yy}}
~,}
\eqn\bbkb{
\rho\equiv \rho_1+i\rho_2=B_{xy}+i\sqrt G
~,}
where $G=G_{xx}G_{yy}-G_{xy}^2$.
The (left,right) scaling dimensions ($h,\bar h$) of a vertex operator
with momenta $n_x, n_y$ and windings $w_x,w_y$ are in this background
given by the following compact formulae (see $e.g.$ \GiveonFU)
\eqn\bbla{
h=\frac{1}{4\rho_2 \tau_2}\bigg| (n_x-\tau n_y)- \rho (w_y+\tau
w_x)\bigg|^2
~,}
\eqn\bblb{
\bar h=\frac{1}{4\rho_2 \tau_2}\bigg| (n_x-\tau n_y)-\bar \rho
(w_y+\tau w_x)\bigg|^2
~.}

For the deformation \baj\ of the diagonal torus we have
\eqn\bbm{
G_{xx}=R_X^2~, ~ ~ G_{yy}=R_Y^2~, ~ ~ G_{xy}=\tilde \lambda_+ R_X R_Y
~, ~ ~ B_{xy}= \tilde \lambda_- R_X R_Y
~, }
or
\eqn\bbn{
\tau=\frac{R_X}{R_Y}\Big(\tilde \lambda_++i\sqrt{1-\tilde
\lambda_+^2}\Big)~, ~ ~
\rho=R_X R_Y\Big(\tilde \lambda_- +i \sqrt{ 1-\tilde \lambda_+^2}\Big)
~.}
$R_X$ is arbitrary and $R_Y=Q$. The latter has been fixed by the GSO
projection at the supersymmetric point. By setting
\eqn\bbo{\eqalign{
&p_n=\frac{n_x}{R_X}~, ~~ p_w= w_x R_X
\cr
&q_n=\frac{n_y}{R_Y}~, ~ ~ q_w= w_y R_Y
}}
we find for the scaling dimensions $(h,\bar h)$ the following
expressions
\eqn\bbpa{\eqalign{
h=&\frac{1}{8(1+ \tilde \lambda_+)}\Big[ p_n+q_n+\big(1+\tilde
\lambda_++\tilde \lambda_-\big)p_w+
\big(1+\tilde \lambda_+-\tilde \lambda_-\big)q_w\Big]^2+
\cr
+&\frac{1}{8(1-\tilde \lambda_+)}\Big[ p_n-q_n+\big(1-\tilde \lambda_+
-\tilde \lambda_-\big)p_w-
\big(1-\tilde \lambda_+ +\tilde \lambda_-\big)q_w\Big]^2
~,}}
\eqn\bbpb{\eqalign{
\bar h=&\frac{1}{8(1+\tilde \lambda_+)}\Big[ p_n+q_n-\big(1+\tilde
\lambda_+-\tilde \lambda_-\big)p_w-
\big(1+\tilde \lambda_++\tilde \lambda_-\big)q_w\Big]^2+
\cr
+&\frac{1}{8(1-\tilde \lambda_+)}\Big[ p_n-q_n-\big(1-\tilde \lambda_+
+\tilde \lambda_-\big)p_w+
\big(1-\tilde \lambda_+ -\tilde \lambda_-\big)q_w\Big]^2
~.}}
It follows that the difference $h - \bar h = p_n p_w + q_n q_w$ is
$\tilde \lambda_{\pm}$-independent, so the physical state condition
$L_0=\bar L_0$ will continue to hold after the deformation, if we impose
it from the beginning.

In general, the scaling dimensions \bbpa, \bbpb\ will become infinite
at the boundary values $\tilde \lambda_+=\pm 1$ and the deformation
terminates there. The existence of a boundary value is an artifact of
the particular parametrization of the moduli space that we are using.
We will encounter these boundary values again
in the next section, where we derive the effect of the asymmetric
deformations on the CHS geometry \baa. The moduli of asymmetric deformations
is summarized in Figure 1.

The lightest modes of the theory have vanishing momentum and winding
in the flat direction $X$. Some of these modes are actually massless
at the supersymmetric point and as an important check of stability we need
to verify how these masses are shifted by the supersymmetry breaking
deformation. Setting $p_n=p_w=0$ in \bbpa, \bbpb\ we find
\eqn\bbq{
h=h_0+\frac{1}{4(1-\tilde \lambda_+^2)}\Big(\tilde \lambda_+ q_n+\tilde
\lambda_- q_w\Big)^2
~,}
where $h_0$ is the undeformed scaling dimension.
Hence, we see that the mass squared of these modes
receives always a non-negative contribution and that no bulk or localized
tachyons are generated after the deformation. It is also interesting to note that
the leading order deformation of the scaling dimensions appears in general at
second order in the deformation parameters and that a positive mass shift
occurs also for the supersymmetric deformations $|\tilde \lambda_+|
=|\tilde \lambda_-|$.

\bigskip
{\vbox{{\epsfxsize=90mm \nobreak \centerline{\epsfbox{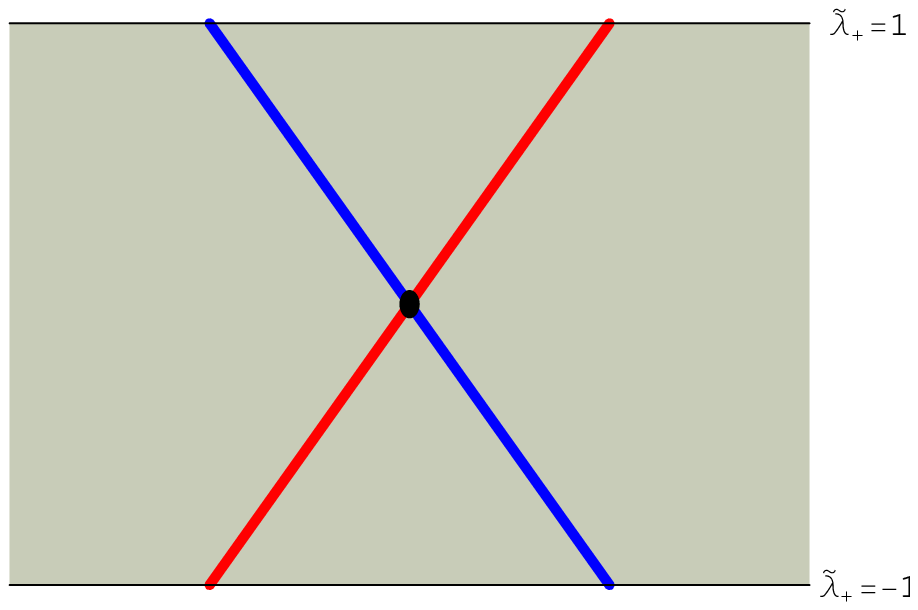}}
\nobreak\bigskip {\raggedright\it \vbox{ {\bf Figure 1.}
 {\it The $(\tilde \lambda_-,\tilde
\lambda_+)$  plane of asymmetric deformations. The shaded region
represents the allowed deformations. The generic point in this
moduli space is a non-supersymmetric theory. The blue and red
lines represent the deformations with $|\tilde \lambda_+|=|\tilde
\lambda_-|$ that preserve half of the spacetime supersymmetry. The
point at the center represents the undeformed type II theory with
sixteen supercharges. } }}}}
\bigskip}

It is instructive to compute the small $\tilde \lambda_{\pm}$
expansion of the scaling dimensions \bbpa, \bbpb\ with generic values for the
$X$ momenta. Expanding the general result up
to second order in the deformation parameters we find
\eqn\bbr{\eqalign{
h&=h_0+\frac{\tilde \lambda_+}{2}\big(p_wq_w-p_nq_n\big)+\frac{\tilde
\lambda_-}{2}\big(p_wq_n-q_wp_n\big)+
\cr
&+\frac{1}{4}\Big[ \tilde \lambda_+^2\big(p_n^2+q_n^2\big)+
\tilde \lambda_-^2 \big(p_w^2+q_w^2\big)-2\tilde \lambda_+\tilde
\lambda_-
(p_w p_n-q_w q_n)\Big]
~.}}
As a further check, we have verified this formula with an explicit
computation of two-point functions in the perturbed theory. In the next section, we
will discuss the effects of the deformation in the low-energy limit of the dual LST
by performing a D5-brane probe analysis in the S-dual of the deformed CHS solution.
We will find good agreement with \bbr, thus achieving a non-trivial test of
holography in a non-supersymmetric context.

As a final comment on the stability of the deformed theories notice
that the masses of the modes with non-vanishing $p_n,p_w$ momenta can be
shifted up or down. This is already clear at the leading linear order in \bbr.
Hence one might worry that under a finite deformation such a mode,
although initially massive, can come down enough to become a tachyon.
For example, this may happen if the difference
\eqn\bbs{
\delta h=h(\tilde \lambda_+,\tilde \lambda_-)\big|_{p_n,p_w}-
h(0,0)\big|_{p_n=p_w=0}
~}
is sufficiently negative. We have verified that $\delta h\geq 0$ for
modes with vanishing winding $p_w=0$. 
In the more general case, we have not been able
to verify conclusively if a tachyon appears after a finite deformation or
not. Certainly, such a tachyon cannot appear for the special deformations
$|\tilde \lambda_+| =|\tilde \lambda_-|$ that preserve
half of the original supersymmetry.

\newsec{Asymmetric deformations of the holographic dual}

As we mentioned earlier, the asymptotically
linear dilaton theories \aab\ appear naturally in the near-horizon
region of NS5-brane configurations and provide
the holographic dual of the non-gravitational Little String Theory
that lives on the fivebranes. For the special case \baa, the appropriate
configuration consists of $k$ coincident parallel NS5-branes that live
deep inside the strongly coupled region of the throat. In the type
IIB case, the low-energy dynamics of this strongly coupled system is
captured by the six-dimensional $\NN=1$ $SU(k)$
super-Yang Mills (SYM) theory that lives on the S-dual configuration of
$k$ parallel D5-branes. As in the usual AdS/CFT correspondence, many aspects of
this holographic relation between bulk and boundary data are known
\refs{\AharonyUB\GiveonPX-\AharonyXN}. There is, in particular,
a well-established dictionary between spacetime chiral primary
operators in the dual low energy SYM theory and corresponding observables in
the non-critical superstring on \baa. In this section, we analyze the
effect of the non-supersymmetric deformation on the dual low energy
gauge theory with a D5-brane probe analysis in the S-dual of the deformed
CHS solution. We work in the large $k$ limit where the spacetime curvatures
are small. For completeness, we give here the relevant part of the type IIB
supergravity action (in the string frame)
\eqn\faa{
\SS_{\rm IIB}= \frac{1}{2 \kappa_{10}^2} \int d^{10}x \sqrt{-g}  \Big[ e^{-
2\Phi}\Big(R+ 4 (\d \Phi)^2
-\frac{1}{12}F^2_{3}  \Big)- \frac{1}{12}G^2_{3} \Big]
}
where $F_{3}$ and $G_{3}$ are the NSNS and RR three-form field
strengths respectively.

\subsec{A D5-brane probe analysis}

The CHS background of $k$ parallel NS5-branes has the form\foot{With
$\alpha'$ reinstated $g_s^2=e^{-2Q\phi/\alpha'}$.}
\eqn\caa{\eqalign{
ds^2&=dx^2_{||}+dx^2+d\phi^2+k(d\theta^2+\sin^2\theta d\phi_1^2
+\cos^2\theta d\phi_2^2)~,
\cr
&B=k\cos^2\theta d\phi_1\wedge d\phi_2 ~, ~ ~ g_s^2=e^{-2Q\phi}
~,}}
where $0\leq \theta \leq \frac{\pi}{2}$, $0\leq \phi_1,\phi_2\leq
2\pi$ and $Q=1/\sqrt k$. $dx^2_{||}$ is the metric element for the five
worldvolume directions $x^{\mu}$, $\mu=0,1,...,4$. We have singled out
the sixth worldvolume direction $x \sim x + 2\pi R_X$, which will take part in the deformation.

With a T-duality transformation along the $\phi_2$ direction we obtain
the equivalent background\foot{For convenience, we have summarized
the general Buscher rules of T-duality in appendix B.}
\eqn\cab{
ds^2=dx^2_{||}+dx^2+d\phi^2+k(d\theta^2+d\phi_1^2)+2d\phi_1 d\phi_2+
\frac{1}{k\cos^2\theta}d\phi_2^2~, ~ ~ B=0
}
and the appropriate dilaton. By defining the new coordinates
\SfetsosXD\
\eqn\cac{
\tilde \phi_1=\phi_1+\frac{\phi_2}{k}~, ~ ~ \tilde
\phi_2=\frac{\phi_2}{k}
}
we can recast \cab\ as
\eqn\cad{
ds^2=dx^2_{||}+dx^2+d\phi^2+k(d\theta^2+d\tilde \phi_1^2+
\tan^2\theta d\tilde \phi_2^2)~, ~ ~
B=0
~.}
Notice that the new variables $(\tilde \phi_1,\tilde \phi_2)$ are
identified under
\eqn\cae{
(\tilde \phi_1,\tilde \phi_2)\sim \Big(\tilde \phi_1+\frac{2\pi}{k},
\tilde \phi_2+\frac{2\pi}{k}\Big)
~.}
The resulting background is none other than
\eqn\caf{
\IR^{4,1}\times S^1_x \times \bigg(S^1_{\tilde \phi_1}
\times \frac{SU(2)_k}{U(1)}\bigg)/\IZ_k
~,}
which is the natural frame for the deformations \baj\ written
in terms of the bosons $X$, $Y$. In terms of the variables $x,\tilde \phi_1$
used above we have the identifications\foot{Note that the $x$ used here is not
the same as the $x$ appearing in \bbj\ .}
\eqn\ident{
x = X  ~, ~ ~ \tilde \phi_1 = \frac{Y}{\sqrt{k}}
~ }

The $\lambda_+ \OO_+ +\lambda_- \OO_-$ deformation
takes the following form in the frame of \cad\foot{In what follows,
we momentarily omit the trivial $dx_{||}^2+d\phi^2$ part of the
metric, which does not participate in the deformation. }
\eqn\cag{\eqalign{
ds^2=&dx^2+k\big(d\theta^2+d\tilde \phi_1^2+\tan^2\theta d\tilde
\phi_2^2\big)
+2 \lambda_+ dxd\tilde \phi_1^2
\cr
&B=\lambda_- ~dx\wedge d\tilde \phi_1
~.}}
Using the identification \ident\ and comparing with \bbm\ we deduce the relation
\eqn\llrel{
\lambda_{\pm} = \sqrt{k} ~ \tilde\lambda_{\pm}
}
between the deformation parameter $\lambda_{\pm}$ used above and
$\tilde \lambda_\pm$ used in section 2.

Going back to the $(\phi_1,\phi_2)$ coordinate system we find
\eqn\cai{\eqalign{
ds^2=&dx^2+k\big(d\theta^2+d\phi_1^2\big)+2d\phi_1d\phi_2+
\frac{1}{k\cos^2\theta}d\phi_2^2+2\lambda_+
dx\Big(d\phi_1+\frac{d\phi_2}{k}\Big)
~,
\cr
&B=\lambda_- ~dx\wedge \Big(d\phi_1+\frac{d\phi_2}{k}\Big)
~.}}
T-dualizing back along $\phi_2$, using the general rules of appendix B,
we get the final form of the deformed near-horizon NS5-brane background
\eqn\cala{\eqalign{
ds^2=& dx^2_{||} + \Big(1+\frac{\lambda^2_--\lambda_+^2}{k}\cos^2\theta\Big)dx^2+d\phi^2 +
k\big(d\theta^2+\sin^2\theta d\phi_1^2+\cos^2\theta d\phi_2^2 \big)+
\cr
&+2\lambda_+ \sin^2\theta dxd\phi_1+2\lambda_- \cos^2\theta dxd\phi_2
~,}}
\eqn\calb{
B=k\cos^2\theta d\phi_1\wedge d\phi_2+\lambda_-\sin^2\theta dx\wedge
d\phi_1
+\lambda_+ \cos^2\theta dx\wedge d\phi_2
~,}
\eqn\calc{
g_s^2=e^{-2Q\phi}\bigg(1-\frac{\lambda_+^2}{k}\bigg)
~.}
This solution, which is exact in the deformation parameters
$\lambda_{\pm}$, exhibits a quadratic contribution to the
$dx^2$ part of the metric and the dilaton. The quadratic contribution
to the metric vanishes when $\lambda_+^2=\lambda_-^2$, which is the case of
the half supersymmetry preserving deformations.
For the non-supersymmetric deformations the
determinant of the metric vanishes identically at the boundary values
$\lambda_+=\pm \sqrt k$. Using the relation \llrel, we see that these
are precisely the boundary values for $\tilde \lambda_+$ that we found
in the previous section, when we computed the scaling dimensions.

The S-dual D5-brane background corresponding to \cala\ - \calc\ reads
\eqn\cama{\eqalign{
ds^2=g_s\bigg[& dx^2_{||} + \Big(1+\frac{\lambda^2_-
-\lambda_+^2}{k}\cos^2\theta\Big)dx^2+ d\phi^2 +
k\big(d\theta^2+\sin^2\theta d\phi_1^2+\cos^2\theta d\phi_2^2 \big)+
\cr
&+2\lambda_+ \sin^2\theta dxd\phi_1+2\lambda_- \cos^2\theta
dxd\phi_2\bigg]
~,}}
\eqn\camb{
G_3=2\cos\theta\sin\theta\big(-k d\phi_1\wedge d\phi_2\wedge d\theta
+\lambda_- dx\wedge d\phi_1\wedge d\theta
-\lambda_+ dx \wedge d\phi_2 \wedge d\theta \big)
~,}
with
\eqn\camc{
g_s^2=e^{2Q\phi}\bigg(1-\frac{\lambda_+^2}{k}\bigg)^{-1}
~}
and $G_3=dC_2$ the RR three-form field strength of the S-dual
$C_2$-field.

The D5-branes couple electrically to the Hodge-dual of $G_3$. This
is a seven-form $G_7=*G_3$ with components
\eqn\can{\eqalign{
&G_{\mu_1\cdots \mu_7}=\sqrt g g^{\nu_1\rho_1} g^{\nu_2\rho_2}
g^{\nu_3\rho_3}
G_{\rho_1\rho_2\rho_3}\epsilon_{\nu_1\nu_2\nu_3\mu_1\cdots \mu_7}=
\cr
&=\sqrt g \Big( g^{\nu_1\phi_1}g^{\nu_2\phi_2}g^{\nu_3\theta}
G_{\phi_1\phi_2\theta}+
g^{\nu_1x}g^{\nu_2\phi_1}g^{\nu_3\theta}G_{x\phi_1\theta}+
g^{\nu_1x}g^{\nu_2\phi_2}g^{\nu_3\theta}G_{x\phi_2\theta}\Big)
\epsilon_{\nu_1\nu_2\nu_3\mu_1\cdots \mu_7}
~.}}
In our case the non-vanishing components of this tensor are
$G_{01234x\phi}$, $G_{01234\phi_1\phi}$, $G_{01234\phi_2\phi}$.
Only the first will play a r\^ole in the DBI analysis and by
straightforward computation we find
\eqn\cao{
G_{01234x \phi }=-2Q
e^{2Q\phi}\Big(1-\frac{\lambda_+^2}{k}\Big)^{-\frac{1}{2}}
.}
The corresponding $C_6$ potential component is
\eqn\cap{
C_{01234x}=-e^{2 Q\phi}\Big(1-\frac{\lambda_+^2}{k}\Big)^{-\frac{1}{2}}
~.}

All the pieces are now in place to proceed with the DBI analysis of a
D5-brane that moves in the background \cama\ - \camc. The DBI+WZ action (up to overall
normalization) of the D5-branes is
\eqn\caq{
\SS_{\rm D5}=\int d^6 \xi ~\frac{1}{g_s}\sqrt{-\det (G_{AB}+F_{AB})}+\int d^6 \xi~ C_6
~,}
where $\xi^A$ ($A,B=0,1,\cdots 5$) are the worldvolume directions,
\eqn\car{
G_{AB}=g_{\mu\nu}\frac{\d X^{\mu}}{\d \xi^A}\frac{\d X^{\nu}}{\d
\xi^B}
}
is the induced metric and $F_{AB}$ the gauge field on the D5-brane.
We will use the static gauge
\eqn\cas{
X^a=\xi^a~, ~ ~ a=0,1,\cdots,4, ~ ~ x=\xi^5
}
and assume that the transverse coordinates $\phi$, $\theta$ are
$\xi$-independent.
The coordinates $\phi_1$ and $\phi_2$ will depend in principle on
$\xi^A$. Hence, the induced metric components are
\eqn\cata{
G_{ab}=g_s\Big(\eta_{ab}+k\sin^2\theta \d_a \phi_1 \d_b\phi_1+
k\cos^2\theta \d_a\phi_2 \d_b \phi_2\Big)
~,}
\eqn\catb{\eqalign{
G_{xx}=g_s\Big(1+\frac{\lambda_-^2-\lambda_+^2}{k}&\cos^2\theta+
k\sin^2\theta \big(\d_x \phi_1\big)^2+k\cos^2\theta\big(\d_x
\phi_2\big)^2+
\cr
+&2\lambda_+\sin^2\theta \d_x\phi_1+2\lambda_- \cos^2\theta
\d_x\phi_2\Big)
~,}}
\eqn\catc{
G_{xa}=g_s\Big(k\sin^2\theta \d_x\phi_1 \d_a \phi_1+
k\cos^2\theta \d_x\phi_2 \d_a \phi_2+\lambda_+ \sin^2\theta
\d_a\phi_1+
\lambda_-\cos^2\theta \d_a\phi_2\Big)
~,}
with $a,b=0,1,\cdots,4$. The WZ term  in \caq\ is
\eqn\cau{
\int d^6 \xi ~C_6=\int d^5 x_{||}dx~ C_{01234x}
~.}

For simplicity, let us set $F_{AB}=0$ and drop the kinetic and
friction terms, $i.e.$ those terms that involve $x^a$-derivatives
$(a=0,1,\cdots,4)$. Then, up to leading order in $\lambda_{\pm}$,
we find the D5-brane Lagrangian
\eqn\cba{
\LL_{\rm D5}=\LL_{\rm D5}\Big|_{\lambda_{\pm}=0}+
\frac{\lambda_+ \sin^2\theta \d_x \phi_1+\lambda_-\cos^2\theta
\d_x\phi_2}
{\sqrt{1+k\sin^2\theta\big(\d_x
\phi_1\big)^2+k\cos^2\theta\big(\d_x\phi_2\big)^2}}
e^{2Q\phi}+\OO(\lambda_{\pm}^2)
~.}

The six-dimensional super Yang-Mills theory that lives on the
D5-branes has four scalar fields in the adjoint of the gauge group. Before the
deformation, the theory has a moduli space parametrized by the vacuum expectation
values of these scalar fields, which simply encode the positions of the
fivebranes in the four transverse directions. Let us combine the four scalar
fields into two complex fields
\eqn\cbb{
A=x_6+ix_7~, ~ ~ B=x_8+ix_9
~.}
In the CHS geometry \caa, $A$ and $B$ take the form
\eqn\cbc{
A=\sqrt k e^{Q\phi} \cos\theta e^{i\phi_2}~, ~ ~
B=\sqrt k e^{Q\phi} \sin\theta e^{i\phi_1}
~.}
With the use of \cbc\ we can recast \cba\ into the more suggestive
form
\eqn\cbd{
\LL_{\rm D5}=\LL_{\rm D5}\Big|_{\lambda_{\pm}=0}
-\frac{i}{k}\frac{\lambda_- A^*\d_x A+\lambda_+ B^*\d_x B}
{\sqrt{1+e^{-2Q\phi}\big(|\d_x A|^2+|\d_x B|^2\big)}} +\OO(\lambda_{\pm}^2)
~.}
Furthermore, at sufficiently low energies $E\ll \frac{1}{R_X}$
we can expand the square root in the denominator in powers
of $\d_x A$ and $\d_x B$. The leading order expansion gives
\eqn\cde{
\LL_{\rm D5} \simeq \LL_{\rm D5}\Big|_{\lambda_{\pm}=0}
-\frac{i}{k}\big(\lambda_- A^*\d_x A+\lambda_+ B^*\d_x B\big)
~.}
As we will show in subsection 3.2, this result can be understood directly from the holographic
prescription. First we show that the expression \cde\ is in perfect agreement with
the perturbative expansion of the scaling dimensions \bbr\ in the bulk.

Using the Fourier expansion of the Higgs fields $A$ and $B$
\eqn\cdf{
A=\sum_{n\in \IZ}A_n e^{-inx/R_{X}}~, ~ ~
B=\sum_{n\in \IZ}B_n e^{-inx/R_{X}}
}
we can rewrite \cde\ in six dimensions as
\eqn\cdg{
\LL_{\rm D5}\simeq \LL_{\rm D5}\Big|_{\lambda_{\pm}=0}
-\frac{1}{k}\sum_{n\in \IZ} \frac{n}{R_X}\big(\lambda_- |A_n|^2+
\lambda_+ |B_n|^2\big)
~.}
Taking into account the normalization of the kinetic
terms, this result yields the mass shifts
\eqn\cdga{
\delta M^2(A_n)=-\frac{2n}{R_X k}\lambda_-~, ~ ~
\delta M^2(B_n)=-\frac{2n}{R_X k}\lambda_+
~.}

At the same time, we know from the bulk calculation \bbr\ that
a general mode with $p_w=0$ and $p_n=\frac{n}{R_X}$
will exhibit the leading order mass shift\foot{We set $p_w=0$ here,
because the DBI action only captures field theory effects.}
\eqn\dci{
\delta M_n^2=-\frac{2n}{R_X}\Big(\tilde \lambda_+ q_n +\tilde \lambda_-
q_w\Big)
~.}
As explained in \AharonyVK\ the $U(1)_A\times U(1)_B$ rotation
symmetries of the $A$, $B$ planes are embedded in the $SU(2)_L\times SU(2)_R$
symmetry of the CHS background in the following way. The generator
of $U(1)_A$ can be taken as $K_3^{(tot)}-\bar K_3^{(tot)}$,
and the generator of $U(1)_B$ as $K_3^{(tot)}+\bar K_3^{(tot)}$.
The corresponding charges can be normalized so that
\eqn\dcj{\eqalign{
&\Big(K_3^{(tot)}+\bar K_3^{(tot)}\Big) (A)=0~, ~ ~
\Big(K_3^{(tot)}-\bar K_3^{(tot)}\Big)(A)=1~,
\cr
&\Big(K_3^{(tot)}+\bar K_3^{(tot)}\Big) (B)=1~, ~ ~
\Big(K_3^{(tot)}-\bar K_3^{(tot)}\Big)(B)=0
~.}}
As a result, single-trace gauge theory operators of the form
$\tr(A_n^k)$, $\tr(B_n^k)$ correspond respectively 
to the string theory winding and momentum $\NN=2$ Liouville-type Kaluza 
Klein vertex operators
\eqn\dcja{
\VV_A=e^{-\varphi-\bar \varphi} e^{i\frac{n}{R_X}(X+\bar X)}
e^{-\frac{1}{Q}(\phi-i(Y+\bar Y))} ~, ~ ~
\VV_B=e^{-\varphi-\bar \varphi} e^{i\frac{n}{R_X}(X+\bar X)}
e^{-\frac{1}{Q}(\phi-i(Y-\bar Y))}
~.
}
Applying the formula \dci\ to these modes and taking into account
the redefinition \llrel\ we find the mass shifts
\eqn\dcjb{
\delta M^2(\VV_A)=-\frac{2n}{R_X k} \lambda_- k~, ~ ~ 
\delta M^2(\VV_B)=-\frac{2n}{R_X k} \lambda_+ k
~,} 
which are fully consistent with the DBI result \cdga.

In addition, the bulk computation of scaling dimensions predicts
a second order mass shift, which is not expected a priori to appear in
the same form in the dual gauge theory. For the lowest Kaluza-Klein
modes ($p_n=p_w=0$) the mass shift predicted by \bbr\ (see also \bbq) is
\eqn\dcl{
\delta M^2=\Big(\tilde \lambda_+ q_n+\tilde \lambda_- q_w\Big)^2
~.}
Applying this formula to the vertex operators \dcja\ (with Kaluza Klein 
momentum $n=0$), we find a result that, up to a factor of $k$, implies in gauge
theory the positive mass squared perturbation
\eqn\dcm{
\frac{\lambda_-^2}{2k^2}|A|^2+\frac{\lambda_+^2}{2k^2}|B|^2
~.}
This result can also be reproduced from the DBI analysis.
Indeed, the potential that follows from the action \caq\ is
\eqn\dcn{\eqalign{
U_{\rm D5}&=e^{2Q\phi}\bigg[\Big(1-\frac{\lambda_+^2}{k}\Big)^{-1}
\sqrt{1+\frac{\lambda_-^2-\lambda_+^2}{k}\cos^2\theta}
-\Big(1-\frac{\lambda_+^2}{k}\Big)^{-\frac{1}{2}}\bigg]
\cr
&=e^{2Q\phi}\bigg[\frac{\lambda_+^2}{2k}\sin^2\theta+
\frac{\lambda_-^2}{2k}\cos^2\theta\bigg]+\OO(\lambda_{\pm}^4)=
\frac{\lambda_-^2}{2k^2}|A|^2+\frac{\lambda_+^2}{2k^2}|B|^2
+\OO(\lambda_{\pm}^4)
~.}}
The zeroth order potential is vanishing as expected by supersymmetry
at the supersymmetric point, but the next order contribution is quadratic
and agrees with the bulk expectation \dcm. As we discuss in the
next subsection, this is a rather non-trivial check of the
holographic duality in this non-supersymmetric context.

\subsec{Discussion}

According to the general bulk-boundary correspondence
the bulk gauge fields $(\AA_{L,R})^a_{\mu}$ (see above \defg)
correspond in the dual non-gravitational theory to a set of global Noether currents. As a result,
the asymmetric bulk deformation \defg\ maps on the gauge theory (fivebrane) side to
the Lagrangian deformation\foot{We thank David Kutasov for pointing this out to us.}
\eqn\llc{
\delta \LL_{\rm D5} \propto \lambda_- ( J_R - J_L) + \lambda_+ (J_R +J_L)
~, }
where $J_{L,R}$ are the $U(1)_{L,R}$ R-symmetry currents\foot{In these
expressions we set the fermions to zero.}
\eqn\jjc{\eqalign{
J_R &= \frac{i}{2} ( A^* \partial_x A - \partial_x A^* A ) +
\frac{i}{2} ( B^* \partial_x B - \partial_x B^* B ) ~ , \cr
J_L & =-\frac{i}{2} ( A^* \partial_x A - \partial_x A^* A ) +
\frac{i}{2} ( B^* \partial_x B - \partial_x B^* B ) ~ .
}}
As stated earlied, \llc, \jjc\ reproduce the last part in \cde.

An alternative way  to understand this result relies on the use of the correspondence
between states in the CHS geometry \baa\ and chiral primary operators in the dual
low energy SYM theory, studied in \refs{\AharonyUB,\AharonyXN}. According to the general
dictionary, the worldsheet vertex operator $\d Y\bar \d Y$ is dual
to the symmetric traceless operator
\eqn\dda{
\tr (A^*A-B^*B)
~.}
When we add this operator to the SYM Lagrangian, one of the complex
Higgs fields becomes massive and the other tachyonic.

One can show that the asymmetric deformations $\int d^2z~ \OO_{\pm}$
belong to the same supersymmetry multiplet as  $\int d^2 z~ \d Y\bar \d Y$.
Indeed, observe that
\eqn\ddb{\eqalign{
Q^2_{\dot +} Q^1_+ \cdot \d Y&=\sqrt{\frac{1}{k}} Q^2_{\dot +} Q^1_+
\cdot \d H_4=
-\frac{i}{2 \sqrt {k}} e^{-\varphi+iH_2}~,~ ~
\cr
Q^1_{\dot +} Q^2_+ \cdot \d Y&=\sqrt{\frac{1}{k}} Q^1_{\dot +} Q^2_+
\cdot \d H_4=
-\frac{i}{2 \sqrt{k}} e^{-\varphi-iH_2}
~.}}
$Q^i_{\alpha}, Q^i_{\dot \alpha}$ ($i=1,2$, $\alpha,\dot \alpha=\pm$)
are spacetime supercharges and $H_2, H_4$ are bosons that bosonize
the appropriate worldsheet fermions (see appendix A for a complete list of our
conventions and the explicit form of the spacetime supercharges).
Therefore, the vertex operator
\eqn\ddba{
e^{-\varphi}\psi^X=\frac{1}{\sqrt
2}e^{-\varphi}\big(e^{iH_2}+e^{-iH_2}\big)
~,}
which is the (-1)-picture version of $\d X$, can be rewritten as
\eqn\ddc{
e^{-\varphi}\psi^X=i\sqrt{2k}(Q^2_{\dot +}Q^1_+ +Q^1_{\dot +}Q^2_+)\cdot
\d Y
~.}
In a similar fashion for the right-movers, one can show that
\eqn\ddd{
e^{-\bar \varphi}\bar \psi^X=i\sqrt{2k}(\bar Q^2_{\dot +}\bar Q^1_+
+\bar Q^1_{\dot +}\bar Q^2_+)\cdot \bar \d Y
~.}
Combining these results and using the equivalence between
the $(-1)$-picture and $0$-picture vertex operators,
we deduce for $\OO_{\pm}$ (see eq.\ \bak)
\eqn\dde{
\OO_{\pm}=i\sqrt{2k}\Big(Q^2_{\dot +}Q^1_+ +Q^1_{\dot +}Q^2_+
\pm(\bar Q^2_{\dot +}\bar Q^1_+ +\bar Q^1_{\dot +}\bar Q^2_+)\Big)
\cdot \d Y \bar \d Y
~.}
Finally, by using the correspondence between the vertex operator
$\d Y \bar \d Y$ and the gauge theory operator \dda\ we find that the
gauge theory dual of $\OO_{\pm}$ is the descendant
\eqn\ddf{
i\Big(Q^2_{\dot +}Q^1_+ +Q^1_{\dot +}Q^2_+
\pm(\bar Q^2_{\dot +}\bar Q^1_+ +\bar Q^1_{\dot +}\bar Q^2_+)\Big)
\tr (A^*A-B^*B)
~.}
Besides terms involving fermions, \ddf\
includes the gauge theory operators
\eqn\ddg{
\tr(A^*\d_x A)~ ~{\rm and} ~ ~ \tr(B^*\d_x B)
~,}
thus verifying the results arising from the DBI analysis
\cde\ and the scaling dimension analysis \bbr\ at leading
order in the deformation parameters $\lambda_{\pm}$.

At second order in the deformation, both the bulk analysis \bbr\ and
the DBI analysis \dcm\ suggest that we should add to the gauge theory
Lagrangian the single trace operator
\eqn\ddi{
\lambda_-^2 \tr(A^*A)+\lambda_+^2\tr(B^*B)
~.}
This operator is symmetric, but not traceless. As a non-chiral primary
operator in a non-supersymmetric theory, it is not a priori expected to
appear simultaneously on both sides of the duality, especially beyond the
leading order of the deformation. Here we find that it does.

\newsec{Supergravity description of rotating fivebranes}

So far we have discussed the effect of the asymmetric deformations
\baj\ on the asymptotic CHS background \baa. Because
of the linear dilaton, this background exhibits a strong coupling
singularity at $\phi\rightarrow -\infty$. This singularity can be resolved by
adding to the worldsheet Lagrangian the $\NN=2$ Liouville interaction,
which in superfield notation is
\eqn\eaa{
\delta \SS=\mu \int d^2 z d^2\theta ~ e^{-\frac{1}{Q}(\phi+iY)}~+~c.c.
}
An alternative way to resolve the singularity is to replace the
$\IR_{\phi}\times S^1$ part of the geometry \bac\ with the $\NN=2$
Kazama-Suzuki supercoset $SL(2)_k/U(1)$ \KazamaQP\ at level
$k=\alpha'/Q^2$. The target space of this conformal
field theory has a cigar-shaped geometry and provides a geometric
cut-off to the strong coupling singularity.

The $\NN=2$ Liouville theory \eaa\ and the $\NN=2$ Kazama-Suzuki model
are known to be equivalent by mirror symmetry \HoriAX. In terms of
NS5-branes both deformations take us into the Higgs branch of the theory, where
the NS5-branes are separated symmetrically along a circle in the
transverse space in an appropriate double scaling limit \GiveonPX.

In the presence of the asymmetric deformations \baj\ the $\NN=2$
Liouville potential \eaa\ is an irrelevant interaction on the worldsheet.
The moduli space of the fivebranes has been lifted and $k$ parallel
NS5-branes at arbitrary positions do not any longer constitute a static
configuration. On the worldsheet, a generic configuration will
be captured by a time-dependent interaction. A natural guess is
\eqn\eab{
\delta \SS=\mu \int d^2 z d^2\theta ~ e^{i\omega
t-\frac{1}{Q}(\phi+iY)}~+~c.c.
}
This is classically marginal when the frequency $\omega$ is
\eqn\eac{
\omega^2 =\frac{1}{Q^2}\frac{\lambda_+^2}{1-\lambda_+^2}
~.}
Interestingly, this expression is $\lambda_-$-independent.
Notice that by defining a new boson $y$ via the relation
\eqn\ead{
\frac{1}{Q}y=-\omega t+\frac{1}{Q}Y
}
we can recast \eab\ as the usual $\NN=2$ Liouville interaction
\eqn\eae{
\delta \SS=\mu \int d^2 z d^2\theta ~ e^{-\frac{1}{Q}(\phi+iy)}~+~c.c.
}
The new boson $y$ is canonically normalized provided that
\eqn\eaf{
\d Y(z) \d Y(0)\sim -\frac{1}{1-\lambda_+^2}\log z
~}
in the presence of the asymmetric deformation \baj.
We have verified this OPE explicitly for the case of generic
$\lambda_+$ and $\lambda_-=0$. Thus, we have a strong indication
that the time-dependent deformation \eab\ is actually exactly marginal.
In spacetime it describes a circular array of fivebranes rotating with
constant angular velocity in the presence of the deformation \baj.

It is interesting to ask if there is a corresponding description of
rotating fivebranes in the language of a deformed $\NN=2$ Kazama-Suzuki
supercoset $SL(2)_k/U(1)$. In the rest of this section, we will address this
problem in the limit of large $k$, where the spacetime curvature is everywhere
small and we can use a supergravity description.

Before the asymmetric deformation \baj\ the Kazama-Suzuki resolution
of the background \bac\ is given by the exact conformal field theory
\eqn\eba{
\IR^{5,1}\times \bigg(\frac{SL(2)_k}{U(1)}\times
\frac{SU(2)_k}{U(1)}\bigg)/\IZ_k
~.}
The string frame metric, $B$-field and dilaton for this solution are
\eqn\ebb{
ds^2=dx_{||}^2+dx^2+k\big(d\rho^2+\tanh^2\rho d\tilde
\phi_1^2+d\theta^2+
\tan^2\theta d\tilde \phi_2^2\big)
~,}
\eqn\ebc{
B=0~, ~ ~ g_s^2=\frac{1}{k \cos^2\theta \cosh^2\rho}
~.}
At $\rho\rightarrow \infty$ this background asymptotes to \cad.
Before proceeding any further, we should raise the following word of
caution. The supergravity solution \ebb, \ebc\ belongs to a well-known list of
examples \refs{\BakasBA\BergshoeffCB-\BakasHC} where the underlying theory is
expected to be supersymmetric, because of extended worldsheet supersymmetry,
but the supergravity solution is manifestly not supersymmetric. For that reason,
it was argued in \IsraelIR\ that the supergravity solution \ebb, \ebc\ is not the
correct low-energy effective description of \eba. The correct description is
the T-dual background, which asymptotes to the CHS solution \baa\
and is manifestly supersymmetric. Having said this, our strategy in the
ensuing will be the following. We start by analyzing the effects of the
asymmetric deformations \baj\ on \ebb, \ebc\ and then after a series of
manipulations we T-dualize to a background that asymptotes to the
deformed CHS solution. As we will see, this approach gives results
which are in qualitative agreement with the picture implied by the
analysis of the previous sections. A similar approach for the case of
the $\int d^2 z~ \d Y\bar \d Y$ deformation was adopted in \ItzhakiZR.

As we showed in the previous section, the asymptotic form of the
$\lambda_+ \OO_+ +\lambda_-\OO_-$ deformation of \ebb, \ebc\ is\foot{In
this section we explicitly include the $(dx^0)^2$ part of the metric in our formulae,
because time will play a crucial r\^ole in what follows. The remaining four worldvolume
coordinates do not participate in our manipulations and will be left implicit.}
\eqn\ebda{
ds^2=-\big(dx^0\big)^2+dx^2+k\big(d\rho^2+d\theta^2+
d\tilde \phi_1^2+\tan^2\theta d\tilde
\phi_2^2\big)+2\lambda_+dxd\tilde\phi_1
~,}
\eqn\ebdb{
B=\lambda_-~ dx\wedge d\tilde\phi_1~, ~ ~
g_s^2=\frac{2}{k\cos^2\theta}e^{-2\rho}\bigg(1-\frac{\lambda_+^2}{k}\bigg)^{\frac{1}{2}}
~.}
For pedagogical reasons, in this section we will focus on the $\OO_-$
deformation, $i.e.$ we will set $\lambda_+=0$ and $\lambda_-=\lambda$. This simple
case is rather instructive and captures the basic features of the generic
situation. More comments on the general $\lambda_{\pm}$ deformation will appear at
the end of this section.

It is convenient to T-dualize the asymptotic solution \ebda, \ebdb\
along the direction $\tilde \phi_1$. For $\lambda_+=0$ we get
\eqn\ebea{
ds^2=-(dx^0)^2+\bigg(1+\frac{\lambda^2}{k}\bigg)dx^2+2\frac{\lambda}{k}dxd\tilde
\phi_1
+k(d\rho^2+d\theta^2+\tan^2\theta d\tilde \phi_2^2)+\frac{1}{k}d\tilde
\phi_1^2
~,}
\eqn\ebeb{
B=0~, ~ ~ g_s^2=\frac{2}{k\cos^2\theta}e^{-2\rho}
~.}
The T-duality has allowed us to convert the non-zero $B$-field to
an off-diagonal component of the metric. Then we can diagonalize the
metric by using the coordinate transformation
\eqn\ebf{
\sqrt{1+\frac{\lambda^2}{k}}x=\frac{1}{\sqrt 2}(X+\Phi)~, ~ ~
\frac{1}{\sqrt k}\tilde \phi_1=\frac{1}{\sqrt 2}(X-\Phi)
~}
to obtain
\eqn\ebg{
\bigg(1+\frac{\lambda^2}{k}\bigg)dx^2+2\frac{\lambda}{k}dxd\tilde
\phi_1
+\frac{1}{k}d\tilde \phi_1^2
=\alpha_+dX^2+\alpha_- d\Phi^2
~}
where we have defined
\eqn\ebi{
\alpha_{\pm} \equiv 1\pm\frac{\lambda}{k}\sqrt
{\frac{k}{1+\frac{\lambda^2}{k}}}
~.}

Finally, it will be useful to define the new time coordinate $t$ as
\eqn\ebj{
t=\frac{1}{\sqrt{\alpha_-}} x^0
~,}
and perform the boost\foot{In \ebg\ $X$ and $\Phi$ appear on equal
footing. Hence, we could equally well perform the boost on the $(x^0,X)$ plane.
The meaning of this choice will be clarified below.}
\eqn\ebk{
\Phi_{\rm new}=C\Phi+St~, ~ ~ t_{\rm new}=Ct+S\Phi~, ~ ~ C^2-S^2=1
~.}
For now the boost parameter $C$ is an arbitrary number.
It will be fixed uniquely in what follows by requiring that we get
a regular background without conical singularities.
In the new coordinates the background \ebea, \ebeb\ takes the form
\eqn\ebla{
ds^2=-\alpha_- dt_{\rm new}^2+\alpha_+ dX^2 +\alpha_- d\Phi_{\rm new}^2+
k(d\rho^2+d\theta^2+\tan^2\theta d\tilde\phi_2^2)
~,}
\eqn\eblb{
B=0~, ~ ~ g_s^2=\frac{2}{k\cos^2\theta}e^{-2\rho}
~.}

We deform this solution in the following way
\eqn\ebma{\eqalign{
ds^2=&-\alpha_- dt_{\rm new}^2+
\frac{1}{2}(\coth^2\rho +1)(\alpha_+ dX^2+\alpha_- d\Phi_{\rm new}^2)-
\cr
&-\frac{\sqrt {\alpha_+\alpha_-}}{\sinh^2\rho}dXd\Phi_{\rm new}+
k(d\rho^2+d\theta^2+\tan^2\theta d\tilde\phi_2^2)
~,}}
\eqn\ebmb{
B=0~, ~ ~ g_s^2=\frac{1}{k\cos^2\theta \sinh^2\rho}
~.}
This background is special, because
\item{($a$)}  It is an exact solution of the lowest order $\alpha'$
equations
of motion. To verify this property perform the change of coordinates
\eqn\ebn{
\sqrt {\alpha_+} X=\frac{1}{\sqrt 2}(y+z)~, ~ ~
\sqrt{\alpha_-}\Phi_{\rm new}=\frac{1}{\sqrt 2}(y-z)
~}
to obtain
\eqn\ebo{\eqalign{
&kd\rho^2+\frac{1}{2}(\coth^2\rho +1)(\alpha_+ dX^2+\alpha_-
d\Phi_{\rm new}^2)
-\frac{\sqrt {\alpha_+\alpha_-}}{\sinh^2\rho}dXd\Phi_{\rm new}=
\cr
&=dy^2+\coth^2\rho dz^2+kd\rho^2
~.}}
This is the well-known trumpet solution, which describes the $U(1)$
vector gauging of $SL(2,\IR)$.
\item{($b$)} At $\rho\rightarrow \infty$ eqs.\ \ebma\ and \ebmb\ reduce
respectively to \ebla, \eblb\ as required.
\item{($c$)} For $\lambda=0$ (and $C=1,S=0$), we will see in a moment
that after the appropriate manipulations \ebma, \ebmb\ give rise to
the anticipated cigar deformation \ebb\ of the asymptotic geometry
\cad.

The next step is to re-express \ebma, \ebmb\ in terms of the original
coordinates $(x,\tilde \phi_1)$ using \ebk, \ebf . With straightforward algebra
we find the metric
\eqn\ebp{\eqalign{
ds^2&=\bigg(1+\frac{\lambda^2}{k}\bigg)
\bigg[1+\frac{(\sqrt{\alpha_+}
-\sqrt{\alpha_-}C)^2}{4\sinh^2\rho}\bigg]dx^2
+\frac{1}{k}\bigg[
1+\frac{(\sqrt{\alpha_+}
+\sqrt{\alpha_-}C)^2}{4\sinh^2\rho}\bigg]d\tilde\phi_1^2
\cr
&+\frac{1}{\sqrt k} \sqrt {1+\frac{\lambda^2}{k}}
\bigg[\alpha_+-\alpha_-+\frac{\alpha_+-\alpha_-C^2}{2\sinh^2\rho}\bigg]dxd\tilde\phi_1
\cr
&+\frac{S}{\sqrt 2 \sinh^2\rho}\sqrt{1+\frac{\lambda^2}{k}}
\Big(C\alpha_--\sqrt{\alpha_+\alpha_-}\Big)dxdt
-\frac{S}{\sqrt {2k}\sinh^2\rho}\Big(
C\alpha_-+\sqrt{\alpha_+\alpha_-}\Big)
dt d\tilde \phi_1
\cr
&- \alpha_-\bigg[1-\frac{S^2}{2\sinh^2\rho}\bigg]dt^2+
k(d\rho^2+d\theta^2+\tan^2\theta d\tilde\phi_2^2)
~.}}
The $B$-field and dilaton are still given by \ebmb. This solution
has a strong curvature and $g_s$ coupling singularity at $\rho=0$,
but soon we will T-dualize to a regular background.

As a trivial check, notice that when $\lambda=0$ (and $C=1,S=0$)
we have $\alpha_+=\alpha_-=1$ and the metric \ebp\ becomes
\eqn\ebq{
ds^2=-dt^2+dx^2+\frac{1}{k}\coth^2\rho d\tilde \phi_1^2+
k(d\rho^2+d\theta^2+\tan^2\theta d\tilde\phi_2^2)
~,}
which can be T-dualized back along $\tilde \phi_1$ to obtain
the cigar geometry
\eqn\ebr{
ds^2=-dt^2+dx^2+k(d\rho^2+\tanh^2\rho d\tilde
\phi_1^2+d\theta^2+\tan^2\theta
d\tilde \phi_2^2)
~.}
This is precisely the deformation we wanted to obtain (see eq.\ \ebb).

Next, we T-dualize \ebp\ along the direction $\tilde \phi_1$.
This gives rise to a background of the form
\eqn\ebs{
ds^2=\sum_{r,s}  \bigg[G_{rs}-\frac{G_{r\tilde \phi_1} G^2_{s\tilde \phi_1}}{G_{\tilde
\phi_1\tilde\phi_1}}\bigg] dx^r dx^s
+\frac{1}{G_{\tilde \phi_1\tilde\phi_1}}d\tilde \phi_1^2
+k(d\rho^2+d\theta^2+\tan^2\theta d\tilde \phi_2^2)
~,}
\eqn\ebt{ B =\sum_r \frac{G_{r\tilde\phi_1}}{G_{\tilde
\phi_1\tilde\phi_1}} dx^r \wedge d \tilde \phi_1
~,}
and the corresponding dilaton  is $g_s^2=\frac{1}{k\cos^2\theta\sinh^2\rho}
\frac{\det G_{\rm new}}{\det G_{\rm old}}$. In these expressions (and below) the indices $r,s$
correspond to $x,t$ and $G_{ij}$ are the metric components of \ebp.
 Note that in comparison to the original cigar background \ebb, \ebc, additional electric
and magnetic field components have been turned on.

We now argue that this background is regular. A possible
singularity can occur at $\rho=0$. Expanding all the components around
$\rho=0$ we find that a possible $\frac{1}{\rho^2}$ divergence of the
coefficients of $dx^2,dt^2,dxdt$ vanishes automatically (for any value of $C$)
and hence that the background is indeed regular. The string coupling $g_s$ is also finite and
bounded from above everywhere. For the metric coefficient of $d\tilde \phi_1^2$ we
find
\eqn\ebv{
\frac{1}{G_{\tilde
\phi_1\tilde\phi_1}}=\frac{4k}{(\sqrt{\alpha_+}+C\sqrt{\alpha_-})^2}\rho^2
+\OO(\rho^4)
~.}
In order to avoid a conical singularity we have to set
\eqn\eby{
\frac{4k}{(\sqrt{\alpha_+}+C\sqrt{\alpha_-})^2}=k
~.}
Using the definitions \ebi , this fixes the boost parameter $C$ to
\eqn\ebz{
C=\frac{2-\sqrt{1+\frac{\lambda}{k}\sqrt{\frac{k}{1+\frac{\lambda^2}{k}}}}}
{\sqrt{1-\frac{\lambda}{k}\sqrt{\frac{k}{1+\frac{\lambda^2}{k}}}}}
~.}
Notice that by definition the boost parameter $C$ has to be greater
than one. This is true for \ebz\ only if $\lambda\geq 0$. If $\lambda \leq 0$,
we simply repeat the above analysis with a boost on the $(t,X)$
(instead of $(t,\Phi)$) plane.

By changing back to the coordinates $(\phi_1,\phi_2)$ and
a further T-duality along $\phi_2$ we can obtain the background
of $\OO_-$-deformed rotating NS5-branes in the language of the original
CHS throat. After the change of coordinates
$\tilde \phi_1=\phi_1+\frac{\phi_2}{k}$, $\tilde
\phi_2=\frac{\phi_2}{k}$ we find
\eqn\eca{\eqalign{
ds^2&= \sum_{r,s} \bigg[G_{rs}-\frac{G_{r\tilde \phi_1} G^2_{s\tilde \phi_1}}{G_{\tilde
\phi_1\tilde\phi_1}}\bigg] dx^r dx^s
+\frac{1}{G_{\tilde \phi_1\tilde\phi_1}}d
\phi_1^2+\frac{2}{kG_{\tilde\phi_1\tilde \phi_1}}d\phi_1d\phi_2
+
\cr
& +k(d\rho^2+d\theta^2)+\bigg(\frac{1}{k^2G_{\tilde
\phi_1\tilde\phi_1}}+
\frac{1}{k} \tan^2\theta \bigg)  d\phi_2^2
~,}}
\eqn\ecb{
B=\sum_r \frac{ G_{r \tilde \phi_1} }{G_{\tilde \phi_1\tilde\phi_1}} dx^r \wedge \Big( d\phi_1+\frac{1}{k}d\phi_2\Big)
~}
with $g_s^2$ unchanged.

A further T-duality transformation along the direction $\phi_2$
gives rise to our final rotating NS5-brane background, with metric
\eqn\ecc{
ds^2 = \sum_{\mu =1,2,3,4} (dx^\mu)^2  + \sum_{\alpha,\beta = x,t,\phi_1, \phi_2} g_{\alpha \beta} dx^\alpha dx^\beta +  k(d\rho^2+d\theta^2)
 ~,}
\eqn\ecca{
g_{rs}=G_{rs}-\frac{G_{r \tilde \phi_1}
G_{s \tilde \phi_1}}{G_{\tilde \phi_1\tilde\phi_1}}
\bigg(1-\frac{1}{{\cal{G}} }\bigg) ~, ~ ~
g_{r \phi_2}=\frac{kG_{r \tilde\phi_1}}{{\cal{G}}}
~,  }
\eqn\ecce{
g_{\phi_1\phi_1}=\frac{1}{G_{\tilde \phi_1\tilde\phi_1}}
\bigg(1- \frac{1}{{\cal{G}}} \bigg) ~, ~  ~
g_{\phi_2 \phi_2} = \frac{k^2 G_{\tilde \phi_1\tilde\phi_1} }{
{\cal{G}} }
~,}
\eqn\eccb{
g_{r \phi_1}=g_{\phi_1\phi_2}=0
~,}
and $B$-field components
\eqn\ecda{
B_{r \phi_1}=\frac{G_{r \tilde \phi_1}}{G_{\tilde
\phi_1\tilde\phi_1}}
\bigg(1-\frac{1}{ {\cal{G}}} \bigg)
~, ~ ~B_{\phi_1\phi_2}=\frac{k}{{\cal{G}}}
~, }
\eqn\ecdb{
B_{rs }=B_{r \phi_2}=0
~.}
In these expressions $G_{ij}$ refer to the metric components of \ebp,
with $\alpha_{\pm}$ given by \ebi\ and $C$ fixed by the regularity
constraint \ebz. The indices $r,s$ correspond to $x,t$
and we have defined the auxiliary function
\eqn\Gdef{
{\cal{G}} \equiv 1+k\tan^2\theta G_{\tilde \phi_1\tilde\phi_1}
~.}
Finally, the string coupling of the resulting background is given by
the formula
\eqn\dilda{
g_s^2 = \frac{k}{ \cos^2 \theta \sinh^2 \rho} \frac{{\rm det} g}{{\rm
det G}}
~,}
where ${\rm det}g$ is the determinant of the $4 \times 4$ part of the
new $g$-metric along the directions $x, \phi_1,t,\phi_2$
and ${\rm det} G$ is the determinant of the
corresponding $4\times 4$ part of the $G$-metric in \ebp\ along the
directions $x, \tilde \phi_1,t,\phi_2$.

We observe that the final background has vanishing metric
component $g_{t\phi_1}$, but non-vanishing $g_{t\phi_2}$.
This suggests (see \cbc ) that the above solution describes a bunch of fivebranes
rotating in the $A$ plane in agreement with the scaling dimension analysis
of section 2 and the DBI effective action analysis of section 3
that suggest a massive deformation for the Higgs field $A$, but nothing
of the sort for $B$. An analogous rotating solution in supergravity was obtained
in \ItzhakiZR\ for the $\int d^2z ~ \d Y\bar \d Y$ deformation
(see (A.27) in \ItzhakiZR). Another interesting property
of the rotating solution in the presence of the $\OO_-$ deformation are
the non-vanishing $g_{xt}$ and $g_{x\phi_2}$ components, which couple
the worldvolume directions of the fivebranes to the transverse
directions.

Analogous results can be obtained with the use of similar methods
(a combination of coordinate transformations, boosts and T-dualities)
for the more general $\lambda_+\OO_+ +\lambda_- O_-$ deformation.
In particular, for the special case of $\lambda_-=0$ and $\lambda_+$
generic we find a regular solution that has vanishing component
$g_{t\phi_2}$ and non-vanishing $g_{t\phi_1}$. Again in agreement with
the analysis of sections 2 and 3, this describes a bunch of fivebranes
rotating in the $B$ plane. For generic non-vanishing deformation parameters
$\lambda_{\pm}$ both $g_{t\phi_1}$ and $g_{t\phi_2}$ are non-vanishing
and the more involved solution describes rotation in both the $A$ and $B$
planes. For example, this is the case of the half-supersymmetry preserving
deformations that have $|\lambda_+|=|\lambda_-|$.

\newsec{Backreaction at one-loop and the cosmological constant}

{} From the above analysis we learn that the worldsheet interactions
\baj\ are exactly marginal deformations of type II string theory on the
CHS solution \baa\ that break, in general, the spacetime supersymmetry
completely. In the large $k$ limit, supergravity is valid and the
corresponding statement is the following. The type II supergravity action \faa\
has a manifold of solutions parametrized by the deformation parameters
$\lambda_{\pm}$. On this manifold the generic solution is non-supersymmetric.
The two special lines defined by the equation $|\lambda_+|=|\lambda_-|$
are the only exception to this statement. On these lines eight supersymmetries
are restored.

In the previous sections, we gave an explicit construction of these
solutions at all orders in the deformation parameters. We also found rotating
geometries where the string coupling $g_s$ is a finite tunable parameter that can
be chosen as small as we like.

Everything in the above discussion has been at tree-level.
Once we break supersymmetry, interesting new effects can arise at-one
loop. Since there is no exact cancellation between bosons and fermions any
longer, the theory will generate a non-vanishing one-loop contribution
that appears in the low-energy effective action \faa\ as a cosmological constant term
$Z(\lambda_-,\lambda_+)$. This term can have a non-trivial effect,
especially in the case of coincident fivebranes, where a strong
coupling singularity develops deep inside the throat. Even then, however,
the effect will be less drastic and under control in the asymptotic
weakly coupled region. In the Higgs branch, where the strong coupling
problem is ameliorated, the one-loop effect is less drastic and under control
everywhere.

It is interesting to investigate this one-loop backreaction
effect with a supergravity analysis in the large $k$ limit. 
In the asymptotic region, which will be the region of interest here,
we expect on general grounds no drastic effects. In particular, no instabilities 
or time-dependence can arise, because the asymptotic spectrum has a mass gap.
Hence, on general grounds one should still expect a manifold
of solutions parametrized by the deformation parameters $\lambda_\pm$.
To determine more concretely the precise effect of the one-loop
contribution, we can use an action of the form
\eqn\fab{
\SS_{\rm IIB}= \frac{1}{2 \kappa_{10}^2} \int d^{10}x \sqrt{-g}
\Big[ e^{- 2\Phi}\Big(R+ 4(\d \Phi)^2 -\frac{1}{12}F^2_{3}  \Big)-
\frac{1}{12}G^2_{3}+Z \Big]
~.}
The induced cosmological constant $Z$ will be a function of the deformation parameters
$\lambda_\pm$, since it comes from the one-loop correction to the
non-supersymmetric solutions.\foot{We anticipate
that $Z(\lambda_-,\lambda_+)$ is an even function of
$\lambda_-$ and $\lambda_+$ because the
solution is invariant under $\lambda_- \rightarrow - \lambda_-$
provided we also transform $\phi_{2} \rightarrow - \phi_{2}$ (or
alternatively $\lambda_+ \rightarrow - \lambda_+$ along with
$\phi_{1} \rightarrow - \phi_{1}$). Moreover, we expect that $Z$
is zero by supersymmetry when $|\lambda_-|=|\lambda_+|$.}
To capture the effect of the one-loop backreaction to leading order
in $g_s \propto e^{-\phi/\sqrt k}$
we will treat $Z$ in what follows as a field independent quantity parameterized
by the deformation parameters $\lambda_\pm$. A more involved treatment is
needed at higher orders in $g_s$.

Solving the equations of motion that follow from \fab\ we
find the one-loop corrected  solution to the deformed background \cala-\calc ,
\eqn\fad{\eqalign{
ds^2 &= \left( 1 + \frac{1}{2} \sqrt{k} \phi Z g_s^2 \right) dx_{||}^2 +
\left( 1 - 3\sqrt k \phi  Zg_s^2 \right) d\phi^2
\cr
& + \left[ 1+ \left( \frac{k}{8}+\frac{3\sqrt{k}+2\phi}{4\sqrt k}(k-\lambda_+^2)\right)Zg_s^2
+ \left( 1+\frac{k}{8}Zg_s^2\right)\frac{\lambda_-^2 - \lambda_+^2}{k} \cos^2 \theta\right]dx^2
\cr
& + k \left( 1 + \frac{k}{8} Z g_s^2 \right)
\left( d\theta^2+\sin^2\theta d\phi_1^2+\cos^2\theta
d\phi_2^2\right)
\cr
& +2\left(1 + \frac{k}{8} Z g_s^2 \right)(
\lambda_+ \sin^2\theta dx d\phi_1 +  \lambda_-
\cos^2\theta dx d\phi_2)
~,}}
with the Kalb-Ramond two-form field and
$g^2_s$ still given by \calb, \calc.

As expected, we notice that this is a time-independent solution with a warping
of the longitudinal world-volume part of the metric. By construction,
the one-loop correction gives a finite shift to the tree-level solution 
that drops exponentially at the asymptotic infinity.

\newsec{More asymmetric deformations}

Up to this point we have been discussing the properties of the
asymmetric deformations \baj\ for the $d=6$ case with
$\MM=SU(2)/U(1)$ in \aab. A few possible extensions of this
discussion are as follows.

One possibility is to consider a more general deformation of the
form
\eqn\gaa{
\lambda_{+;\mu\nu}\int d^2 z ~ \OO^{\mu\nu}_+ +
\lambda_{-;\mu\nu}\int d^2 z ~ \OO^{\mu\nu}_-
~,}
where
\eqn\gab{
\OO^{\mu\nu}_{\pm}=\d x^{\mu} \bar \d Y\pm \d Y \bar \d x^{\nu}
~, ~ ~ \mu,\nu=0,1,\cdots, 5
~.}
We do not expect new physics from these deformations, since
they can be viewed as a combination of \baj\ and the $SO(5,1)$
Lorentz symmetry group of the theory.

Various aspects of our discussion can be generalized easily to the
other non-critical superstrings in \aab\ with dimensions $d=2,4$ and
general compact manifolds $\MM$. In particular, the whole discussion of
scaling dimensions in section 2 is universal and does not depend on the
details of the theory. Eqs.\ \bbpa\ - \bbr\ will always be true and our
conclusions for the stability of the corresponding non-supersymmetric deformations
will remain unaltered. Certain details of the DBI analysis are case
specific, but the main conclusions about the agreement between
bulk and boundary should of course remain unchanged.
It is also possible to obtain rotating supergravity solutions similar
to those of section 4 as long as we can take an appropriate supergravity limit
in the non-critical superstring.

A potentially more interesting case is the one analyzed recently
in \ItzhakiTU, where \aaa\ reads
\eqn\gag{
\IR^{2,1}\times \IR_{\phi}\times SU(2)_{k_1} \times SU(2)_{k_2}
~.}
This theory, which appears in the near horizon geometry of $k_1$ and
$k_2$ NS5-branes with a three dimensional intersection, exhibits a
mass gap \refs{\ItzhakiTU,\LinNH} and possesses a supergravity limit.

Finally, in this paper we have explored the moduli space of asymmetric
deformations (see figure 1) around the supersymmetric type IIA and type
IIB theories of references \refs{\KutasovUA,\KutasovPV}. In a similar fashion, we
may consider asymmetric deformations around any point in the moduli space presented
in \ItzhakiZR, including for example the type 0A and type 0B theories.
The generic point in the overall moduli space is a non-supersymmetric
theory with stability properties that depend crucially on the details
of the theory.

\newsec{Summary and interesting prospects}

Spacetime supersymmetry can be broken continuously in
non-critical superstring theories with appropriate current-current
deformations on the worldsheet. The deformed theory exhibits a variety of
interesting features: bulk or localized tachyons, a non-supersymmetric
spectrum, time dependence, a tunable cosmological constant.
In this paper we analyzed a set of asymmetric current-current deformations
and showed that the lightest modes receive a non-negative mass squared shift,
which precludes the generation of bulk or localized
tachyons. In a six dimensional example, we verified this effect at
leading order in the deformation by analyzing the deformation on
the dual low-energy SYM theory. For lowest Kaluza-Klein modes
the leading order effect appears at second order and involves a
non-chiral quadratic operator of the Higgs fields.

As another consequence of the mass shifts, we found that the $\NN=2$
Liouville interaction becomes irrelevant on the worldsheet.
The exactly marginal interaction is time dependent and describes
a configuration of rotating NS5-branes. We analyzed this effect
in the supergravity limit and found a manifold of rotating
supergravity solutions at weak $g_s$ coupling.

There is a number of interesting extensions of the results presented
in this paper. First of all, it would be interesting to repeat the
analysis of the $d=6$, $\MM=SU(2)/U(1)$ case for the type IIA case.
The results of section 2 are similar for the type IIA case, but the details of
holography are different. The type IIA low-energy holographic dual is
the six-dimensional $(0,2)$ SCFT and the strong coupling singularity of \baa\
is better studied in eleven dimensional M-theory. It would be interesting to verify
certain statements about holography in this context. Another interesting
direction is to study the effects of the asymmetric deformations in heterotic non-critical
superstring theory (for recent work in this theory see \MurthyEG).

There is also a rich story of open string dynamics on spaces of the form
\eqn\iaa{
\IR^{d-1,1}\times \bigg(\frac{SL(2)_k}{U(1)} \times \MM\bigg)/\Gamma
~.}
D-branes on such spaces have been analyzed in
a series of recent papers (see \IsraelFN\ and references therein)
and are expected to play a key r\^ole in uncovering the inner
workings of LST's and the corresponding holographic dualities.
The dynamics of these branes are also interesting from the gauge theory point of
view. In recent work \refs{\FotopoulosCN,\AshokPY}
it has been verified explicitly that there are appropriate D-brane
configurations on \iaa\ that realize gauge theories with minimal
supersymmetry. For instance, one can obtain four-dimensional $\NN=1$ SQCD
with appropriate D-branes on
\eqn\iab{
\IR^{3,1}\times \frac{SL(2)_1}{U(1)}
~.}
Further work on gauge theories in this general context can be found
in \IsraelZP.

It will be interesting to study the effect of the asymmetric deformations
on D-branes on \iaa. This will give further information about the
dynamics of the non-critical string on \iaa, and will, in particular, clarify
how the breaking of supersymmetry in the bulk affects the dynamics
of the gauge theory on the D-branes. For that purpose, it will be useful to
obtain a better grasp on the exact CFT properties of the asymmetric
deformations on \iaa\ beyond the supergravity limit.

Since we break supersymmetry in a well controlled stringy
environment it is tempting to ask if there are any potential
phenomenological applications of our work. Perhaps this could be
achieved along the lines of \AntoniadisSW\ and would be worth investigating
further. A related question, which is also interesting on its own,
has to do with the higher-loop backreaction problem. As a starting point
in this direction, we analyzed in section 5 the leading order effects
of the one-loop cosmological constant on the tree-level deformed CHS solutions.
In the asymptotic region, where the coupling constant
goes continuously to zero we determined the exact, finite, time-independent shift 
of the one-loop contribution to the tree-level solutions.
It would be interesting to go further and give a more complete
treatment of the backreaction problem in this setting.

Finally, it could be interesting to use similar techniques to further
examine the thermodynamics of near-extremal NS5-brane backgrounds 
and their application to LST at finite temperature. It was recently shown in
\HarmarkDT\  that, in the canonical ensemble, the usual near-extremal NS5-brane 
background is subdominant to a new stable phase of near-extremal M5-branes localized on a transverse circle, having a limiting temperature that lies above the Hagedorn temperature. It is conceivable that there are other stable phases that are relevant to the thermodynamic behavior
of LST.

\vskip10mm
\centerline{\bf Acknowledgments}
\vskip0.2cm

We would like to thank Shinji Hirano, Dan Israel, Elias Kiritsis and
Nikolaos Prezas for useful discussions. Special thanks to David Kutasov
for useful comments on the manuscript and helpful correspondence.
Also thanks to the referee of Nuclear Physics B for his/her comments and for 
pointing out an erroneous statement in the original version of this paper. 
Work partially supported by the European Community's Human Potential
Programme under contract MRTN-CT-2004-005104 `Constituents,
fundamental forces and symmetries of the universe'.

\appendix{A}{Summary of conventions}

In this appendix we summarize the basic conventions used in the
main text. For $k$ parallel NS5-branes wrapped on a circle $S^1$ the
near horizon geometry takes the form
\eqn\appaa{
\IR^{4,1}\times S^1_X\times \IR_{\phi}\times SU(2)_k
~.}
The worldsheet theory on \appaa\ comprises of six free scalars
$x^{\mu}$ $(\mu=0,1,\cdots,5$), six corresponding
free real fermions $\psi^{\mu}$, one linear dilaton scalar
$\phi$ with linear dilaton slope $Q=1/\sqrt{k}$ (we set $\alpha'=1$), the corresponding
free fermion $\psi^{\phi}$ and finally the supersymmetric $SU(2)$ WZW
model at level $k$. In the main text, we compactify the sixth direction
and denote the corresponding boson and fermion as $X$ and $\psi^X$
respectively. The supersymmetric $SU(2)_k$ WZW model comprises of a
bosonic $SU(2)_{k-2}$ WZW model at level $k-2$ and three real free
fermions $\chi^{\pm}$, $\chi^3$. We denote the three (left-moving) $SU(2)$
currents as $K^{\pm}$, $K^3$.

It is convenient to bosonize the above fermions in the following
manner
\eqn\appaba{\eqalign{
&\frac{1}{\sqrt 2}(\psi^1\pm \psi^0)=e^{\pm i H_0}~, ~ ~
\frac{1}{\sqrt 2}(\psi^2\pm i \psi^3)=e^{\pm i H_1}
\cr
&\frac{1}{\sqrt 2}(\psi^X \pm i \psi^4)=e^{\pm i H_2}~, ~ ~
\frac{1}{\sqrt 2}(\psi^{\phi} \pm \chi^3)=e^{\pm i H_3}
~,}}
and
\eqn\appabb{
\chi^{\pm}=e^{\pm i H_4}
~.}
We also define and bosonize the total $K_3^{(tot)}$ current as
\eqn\appac{
K_3^{(tot)}=K^3+\chi^+\chi^-=K^3+i\d H_4=i\sqrt{k}\d Y
~.}
Notice that for $k=2$ the bosonic $SU(2)_{k-2}$ WZW model becomes
trivial and the boson $H_4$ equals $\sqrt 2 Y$.

The above theory has $\NN=(4,4)$ worldsheet supersymmetry.
The construction of the spacetime supercharges makes use of
the $U(1)_R$ current of the $\NN=(2,2)$ worldsheet supersymmetry,
which in our case has the (left-moving) fermionic generators
\eqn\appad{
G^{\pm}=\sqrt 2 \sum_{i=0}^2 e^{\mp i H_i}\d x^i+
Q \chi^{\mp}K^{\pm}
+i \Big( e^{\mp i H_3}(\d \phi\pm i \d Y)-Q\d e^{\mp i H_3}\Big)
~}
and the $U(1)_R$ current
\eqn\appae{
J=-i\d H_4+i\d H_3
~.}
The three complex bosons $x^i$ are defined as
\eqn\appaf{
x^0=\frac{1}{\sqrt 2}(x^0+x^1)~, ~ ~
x^1=\frac{1}{\sqrt 2}(x^2+ix^3)~, ~ ~
x=\frac{1}{\sqrt 2}(X+ix^4)
~.}

Type IIB string theory on \appaa\ exhibits sixteen supercharges
in the
\eqn\appag{
({\bf 4},{\bf 2})+({\bf 4}',{\bf 2}')
}
representations of $SO(5,1)\times SO(4)$. The eight supercharges
$Q^{\pm}_A$ in $({\bf 4},{\bf 2})$ arise from the left-moving sector
of the string and the other eight supercharges $\bar Q^{\pm}_{\dot A}$ in
$({\bf 4}',{\bf 2}')$ arise from the right-moving sector.
In worldsheet terms these supercharges can be written as
\eqn\appaja{
Q^{\pm}_A=\oint \frac{dz}{2\pi i}~
e^{-\frac{\varphi}{2}\mp\frac{i}{2}(H_3-H_4)}\SS_A
~,}
\eqn\appajb{
\bar Q^{\pm}_{\dot A}=\oint \frac{d\bar z}{2\pi i}~
e^{-\frac{\bar \varphi}{2}\mp\frac{i}{2}(\bar H_3-\bar H_4)} \bar
\SS_{\dot A}
~,}
where $\SS_A$, $\bar \SS_{\dot A}$ are the spin fields
\eqn\appaka{
\SS_A=e^{\frac{i}{2}(\alpha_0 H_0+\alpha_1 H_1+\alpha_2 H_2)}
~, ~ ~ \alpha_i=\pm 1~, ~ {\rm even ~ number~ of ~} -'s
~,}
and
\eqn\appakb{
\bar \SS_{\dot A}=e^{\frac{i}{2}(\bar \alpha_0 \bar H_0+\bar \alpha_1
\bar H_1+\bar \alpha_2
\bar H_2)}
~, ~ ~ \bar \alpha_i=\pm 1~, ~ {\rm odd ~ number~ of ~} -'s
~.}

{} From the five dimensional point of view the spacetime supercharges
take the following form. Again, eight supercharges arise from the
left-moving sector
\eqn\appala{
Q^1_{\alpha}=\oint \frac{dz}{2\pi i}~
e^{-\frac{\varphi}{2}+\frac{i}{2}(H_2+H_3-H_4)
+i\frac{\alpha}{2}(H_0+H_1)}~, ~ ~
Q^1_{\dot{\alpha}}=\oint \frac{dz}{2\pi i}~
e^{-\frac{\varphi}{2}-\frac{i}{2}(H_2+H_3-H_4)
+i\frac{\dot{\alpha}}{2}(H_0-H_1)}
~,}
\eqn\appalb{
Q^2_{\alpha}=\oint \frac{dz}{2\pi i}~
e^{-\frac{\varphi}{2}+\frac{i}{2}(-H_2+H_3-H_4)
+i\frac{\alpha}{2}(-H_0+H_1)}~, ~ ~
Q^2_{\dot{\alpha}}=\oint \frac{dz}{2\pi i}~
e^{-\frac{\varphi}{2}+\frac{i}{2}(H_2-H_3+H_4)
-i\frac{\dot{\alpha}}{2}(H_0+H_1)}
~,}
and eight more $\bar Q^i_{\alpha}$, $\bar Q^i_{\dot{\alpha}}$
($i=1,2$) arise from the right-moving sector. In type IIB the latter
are
\eqn\appama{
\bar Q^1_{\alpha}=\oint \frac{d\bar z}{2\pi i}~
e^{-\frac{\bar \varphi}{2}+\frac{i}{2}(\bar H_2+\bar H_3-\bar H_4)
+i\frac{\alpha}{2}(\bar H_0-\bar H_1)}~, ~ ~
\bar Q^1_{\dot{\alpha}}=\oint \frac{d\bar z}{2\pi i}~
e^{-\frac{\bar \varphi}{2}-\frac{i}{2}(\bar H_2+\bar H_3-\bar H_4)
-i\frac{\dot{\alpha}}{2}(\bar H_0+\bar H_1)}
~,}
\eqn\appamb{
\bar Q^2_{\alpha}=\oint \frac{d\bar z}{2\pi i}~
e^{-\frac{\bar \varphi}{2}+\frac{i}{2}(-\bar H_2+\bar H_3-\bar H_4)
+i\frac{\alpha}{2}(\bar H_0+\bar H_1)}~, ~ ~
\bar Q^2_{\dot{\alpha}}=\oint \frac{d\bar z}{2\pi i}~
e^{-\frac{\bar \varphi}{2}+\frac{i}{2}(\bar H_2-\bar H_3+\bar H_4)
-i\frac{\dot{\alpha}}{2}(\bar H_0-\bar H_1)}
~.}
The indices $\alpha,\dot{\alpha}$ take the values $\pm$.
The reduction from six to five dimensions works as follows
\eqn\appana{\eqalign{
&Q^+_A \rightarrow \{Q^1_{\dot{\alpha}},Q^2_{\dot {\alpha}}\}
~, ~ ~
Q^-_A \rightarrow \{Q^1_{\alpha},Q^2_{\alpha}\}
~,
\cr
&\bar Q^+_{\dot A} \rightarrow \{\bar Q^1_{\dot \alpha},\bar Q^2_{\dot
\alpha}\}
~, ~ ~
\bar Q^-_{\dot A} \rightarrow \{\bar Q^1_{\alpha},\bar Q^2_{\alpha}\}
~.}}

\appendix{B}{Summary of the Buscher rules}

For quick reference, we briefly recall here the Buscher rules of
T-duality (see for instance \GiveonFU).
For a general background that has an isometry along the direction
$x^0$, metric of the form ($i,j,...\neq 0$)
\eqn\caj{
ds^2=G_{00}\big(dx^0\big)^2+2G_{0i}dx^0 dx^i+2G_{ij}dx^idx^j
~}
and $B$-field components $B_{0i}$, $B_{ij}$ the T-duality
transformation
along $x^0$ acts in the following manner
\eqn\caka{
G'_{00}=\frac{1}{G_{00}}~, ~ ~ G'_{0i}=\frac{B_{i0}}{G_{00}}~, ~ ~
G'_{ij}=G_{ij}-\frac{G_{i0}G_{0j}-B_{0i}B_{0j}}{G_{00}}
~,}
\eqn\cakb{
B'_{i0}=\frac{G_{0i}}{G_{00}}~, ~ ~
B'_{ij}=B_{ij}-\frac{G_{0i}B_{0j}-B_{0i}G_{0j}}{G_{00}}
~.}
The dilaton transforms as
\eqn\cakc{
g_{s,{\rm new}}^2=g_{s,{\rm old}}^2\frac{\det G_{\rm new}}{\det G_{\rm
old}}
~.}

\listrefs
\bye